\documentclass[sigconf]{acmart}
\usepackage{algorithmic}
\usepackage{caption}
\usepackage[ruled,vlined]{algorithm2e}
\usepackage{booktabs}
\usepackage{pifont}
\usepackage{fancyhdr} % 导入fancyhdr包
\usepackage{color}
\usepackage{soul, framed}
\usepackage{makecell}
\usepackage{xcolor}
\usepackage{colortbl}
\usepackage{graphicx}
\usepackage{textcomp}
\usepackage{xcolor,colortbl}
\usepackage{multirow}
\usepackage{bm,dsfont}
\usepackage{subfigure}
\usepackage{balance}
\usepackage{stmaryrd}
\usepackage{autobreak}
\usepackage{enumitem}
\usepackage{bbding}
\usepackage{pifont}
\usepackage{wasysym}
\newcommand\vldbavailabilityurl{URL_TO_YOUR_ARTIFACTS}
\newcommand\vldbpagestyle{plain} 

\newtheorem{definition}{Definition}
\newtheorem{lemma}{Lemma}

\newtheorem{example}{Example}

\newcommand{\framework}{\textit{TransZero}}

\newcommand{\transformername}{\textit{CSGphormer}}
\newcommand{\searchname}{\textit{IESG}}
\newcommand{\localsearch}{\textit{Local Search}}
\newcommand{\globalsearch}{\textit{Global Search}}

\begin{document}

\title{Efficient Unsupervised Community Search with Pre-trained Graph Transformer}

% \author{Ben Trovato}
% \affiliation{%
%   \institution{Institute for Clarity in Documentation}
%   \streetaddress{P.O. Box 1212}
%   \city{Dublin}
%   \state{Ireland}
%   \postcode{43017-6221}
% }

\author{Jianwei Wang$^{1}$ ~~ Kai Wang$^{2}$ ~~ Xuemin Lin$^{2}$~~Wenjie Zhang$^{1}$ ~~Ying Zhang$^{3}$}
\affiliation{\normalsize{$^{1}${The University of New South Wales, Sydney, Australia}}\\
 \normalsize{$^{2}${Antai College of Economics \& Management, Shanghai Jiao Tong University, Shanghai, China}} \\
 \normalsize{$^{3}${Zhejiang Gongshang University, Hangzhou, China}}}
\email{{jianwei.wang1, wenjie.zhang}@unsw.edu.au, {w.kai, xuemin.lin}@sjtu.edu.cn, ying.zhang@zjgsu.edu.cn}
% \email{jianwei.wang1@unsw.edu.au, w.kai@sjtu.edu.cn, xuemin.lin@sjtu.edu.cn, zhangw@cse.unsw.edu.au, ying.zhang@uts.edu.au}
% \author{Jianwei Wang$^{1}$ ~~ Kai Wang$^{2}$ ~~ Xuemin Lin$^{2}$~~Wenjie Zhang $^{1}$ ~~Ying Zhang$^{3}$ ~~\\
% %\and
% %\alignauthor
%  \normalsize{$^{1}${The University of New South Wales, Sydney, Australia}}\\
%  \normalsize{$^{2}${Shanghai Jiao Tong University, Shanghai, China}} \\
%  \normalsize{$^{3}${University of Technology Sydney, Sydney, Australia}} \\
% \normalsize{jianwei.wang1@unsw.edu.au, kai.wang@sjtu.edu.cn, xuemin.lin@sjtu.edu.cn, zhangw@cse.unsw.edu.au, ying.zhang@uts.edu.au}\\
% }

% \renewcommand{\shortauthors}{Jianwe Wang et al.}
% \renewcommand{\shorttitle}{Research Data Management Track Paper}

%%
%% The abstract is a short summary of the work to be presented in the
%% article.

\begin{abstract}
% In this way, we can encode the community information and graph topology within the pre-trained latent representation without the usage of ground-truth labels. 
% to 1) train the neural models for community score learning, and 2) select the optimal parameters for community identification. property of community 
% , and learning-based model that does not need labels remains under-explored
% finding a connected subgraph that contains the query nodes while having the maximum expected score gain (\searchname). 
Community search has aroused widespread interest in the past decades. 
Among existing solutions, the learning-based models exhibit outstanding performance in terms of accuracy by leveraging labels to 1) train the model for community score learning, and 2) select the optimal threshold for community identification. However, labeled data are not always available in real-world scenarios. To address this notable limitation of learning-based models, we propose a pre-trained graph  \textbf{\underline{Trans}}former based community search framework that uses \textbf{\underline{Zero}} label (i.e., unsupervised), termed \framework. \framework~has two key phases, i.e., the offline pre-training phase and the online search phase. 
Specifically, in the offline pre-training phase, we design an efficient and effective community search graph transformer (\transformername) to learn node representation. 
To pre-train \transformername~without the usage of labels, we introduce two self-supervised losses, i.e., personalization loss and link loss, motivated by the inherent uniqueness of node and graph topology, respectively. 
In the online search phase, with the representation learned by the pre-trained \transformername, we compute the community score without using labels by measuring the similarity of representations between the query nodes and the nodes in the graph.
To free the framework from the usage of a label-based threshold, we define a new function named expected score gain to guide the community identification process.
Furthermore, we propose two efficient and effective algorithms for the community identification process that run without the usage of labels.
Extensive experiments over 10 public datasets illustrate the superior performance of \framework~regarding both accuracy and efficiency.

% Regarding accuracy,  \framework~ even outperforms the semi-supervised state-of-the-art and the supervised state-of-the-art with an average F1-score improvement of 10.01\% and 5.91\%, respectively.
% Regarding efficiency, \framework~achieves a speedup up to 486.07$\times$ and 235.83$\times$ in the offline training phase, and achieves a speedup up to 20.41$\times$ and 56.48$\times$ in the online search phase compared to the semi-supervised state-of-the-art and the supervised state-of-the-art, respectively. 
\end{abstract}


 \maketitle

\pagestyle{\vldbpagestyle}
\ifdefempty{\vldbavailabilityurl}{}{

\begingroup\small\noindent\raggedright\textbf{PVLDB Artifact Availability:}\\
The source code, data, and/or other artifacts have been made available at \url{https://github.com/guaiyoui/TransZero}.
\endgroup
}

% \vspace{-3mm}
\section{Introduction}
\label{sec:Introduction}

\begin{table}

 \centering \scalebox{0.82}{
\begin{tabular}
{@{}lcccc@{}}
\toprule[1.2pt]
Methods  & \makecell[c]{Label \\ Free?}   & \makecell[c]{Structure \\ Flexibility?}    & \makecell[c]{Backbone \\ Model}  & \makecell[c]{Loss \\ Function}          \\ \midrule
CST~\cite{cui2014local} & \Checkmark     & \XSolidBrush &     $k$-core    & -          \\
EquiTruss ~\cite{akbas2017truss}   & \Checkmark      & \XSolidBrush & $k$-truss         & -         \\
M$k$ECS~\cite{DBLP:conf/cikm/AkibaIY13}   & \Checkmark   & \XSolidBrush & $k$-ECC         & - \\
CTC~\cite{huang2015approximate}   & \Checkmark    & \XSolidBrush & $k$-truss & -         \\
QD-GNN~\cite{jiang2022query}   & \XSolidBrush   & \Checkmark & GNN       & Binary Cross Entropy          \\
COCLEP~\cite{li2023coclep} & \XSolidBrush    & \Checkmark & GNN & Contrastive Loss    \\
 
\midrule
\framework~(our)  & \Checkmark & \Checkmark & \makecell[c]{Graph \\ Transformer} &  \makecell[c]{ Contrastive Loss \\ \&  Generative Loss} \\ \bottomrule
\end{tabular}}
 % \begin{flushleft}
 %        \footnotesize $(1):$ We report preparation time $+$ train (query) time; $(2):$ --- indicates out of memory or not finished within 7 days; $(3):$  *** indicates this cell not applicable to this model. 
 %    \end{flushleft}
 \vspace{1mm}
 % Technical
 \caption{Characteristics comparison among CS methods}
 \vspace{-10mm}
 \label{tab:intro_comparasion}
\end{table}

{\color{black}Graphs} play a prominent role in modeling relationships between entities in a system and are applied across diverse domains such as social networks~\cite{tang2010graph, cattuto2013time}, biology networks~\cite{sankar2022sitemotif, wang2020gcncda} and finance networks~\cite{chen2021novel,yang2021financial,bi2022company}. 
{\color{black}As a fundamental problem in graph analytics, community search (CS)~\cite{fang2020survey} has aroused widespread interest in the past decades. Given a set of query nodes, CS aims to find a query-dependent subgraph}, with the resultant subgraph, also referred to as a community, manifesting as a densely intra-connected structure. {\color{black}CS is also relevant and widely applied for tasks in real-world applications}, such as friend recommendation in social networks~\cite{cheng2011personalized, chen2008combinational}, fraud detection in e-commerce platforms~\cite{li2021happens, zhang2017hidden} and protein complex identiﬁcation~\cite{qi2008protein, fang2020survey}.  {\color{black}Given the importance and widespread applications of CS, a set of algorithms are proposed, which include the traditional CS algorithms~\cite{cui2014local, sozio2010community,huang2014querying, akbas2017truss, chang2015index, hu2016querying} and learning-based CS models~\cite{gao2021ics,li2023coclep,jiang2022query}.}

\begin{figure*}
  \centering
  \includegraphics[width=0.90\linewidth]{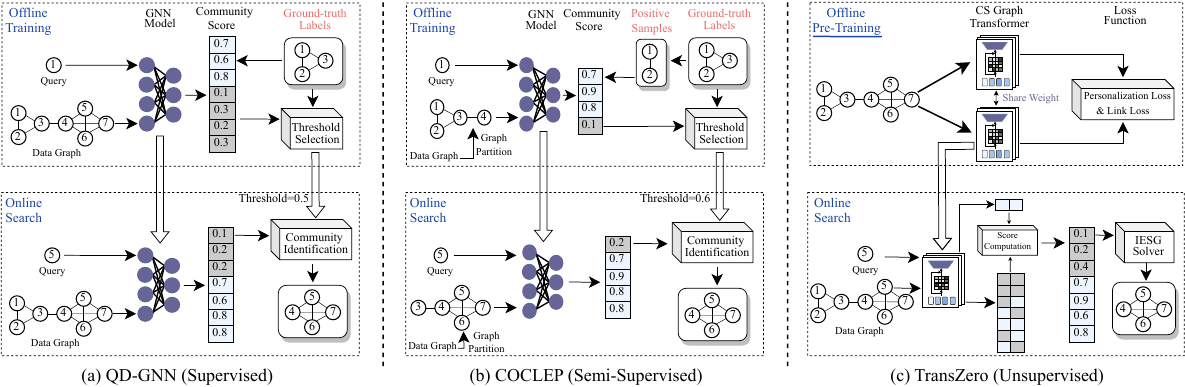}
  \vspace{-5mm}
  \caption{Framework comparisons of learning-based methods for CS }
  \label{fig:frameworks_comparison}
\vspace{-6mm}
\end{figure*}

{\color{black}As summarized in Table~\ref{tab:intro_comparasion} and in the recent survey paper~\cite{fang2020survey}, most of existing traditional CS algorithms characterize the community structure by specific subgraph cohesiveness models such as \textit{k}-core~\cite{cui2014local, sozio2010community}, \textit{k}-truss~\cite{huang2014querying, akbas2017truss, huang2015approximate} and \textit{k}-edge connected component (\textit{k}-ECC)~\cite{chang2015index, hu2016querying, DBLP:conf/cikm/AkibaIY13}, and thus suffer from the limitation known as \textit{structure inflexibility}~\cite{jiang2022query,gao2021ics,li2023coclep}. These fixed subgraph models impose rigid constraints on the topological structure of communities, making it difficult for real-world communities to meet such inflexible constraints. For example, methods based on \textit{k}-core require every node in the found community to have a degree larger than or equal to \textit{k}, which may not be met by real-world communities, particularly for nodes located at the boundary of the community.}

{\color{black}Recently, learning-based approaches such as QD-GNN~\cite{jiang2022query} and COCLEP~\cite{li2023coclep} are emerging in this field due to their outstanding performance in terms of accuracy.} 
As illustrated in Figure~\ref{fig:frameworks_comparison}(a) and Figure~\ref{fig:frameworks_comparison}(b), QD-GNN trains a {\color{black}feature-aggregation model} in a supervised manner where all nodes from the ground-truth community are used for training, and COCLEP trains a {\color{black}feature-aggregation model} in a semi-supervised manner where only a subset nodes from the ground-truth community is used for training.
{\color{black}By incorporating meticulous framework design and leveraging ground-truth information, both QD-GNN and COCLEP can effectively alleviate the limitation of \textit{structure inflexibility} encountered by traditional CS algorithms, and thus demonstrate state-of-the-art performance in the supervised and semi-supervised settings, respectively.}
In a nutshell, they both employ a two-stage framework, consisting of an offline training phase and an online search phase, and rely on the labels to 1) train the neural network for \textit{community score learning} where the community score reflects the membership of the corresponding node w.r.t. the query, and 2) select the optimal threshold from the labeled validation set for \textit{community identification}.

\vspace{1mm}
\noindent \textbf{Motivations.}
While learning-based approaches demonstrate phenomenal performance, a notable limitation of current learning-based methods is their dependence on ground-truth communities. Ground-truth communities are often unavailable or of low quality in real-world scenarios. Additionally, the excessive dependence on ground-truth communities makes it challenging for QD-GNN to generalize and predict unseen communities, as evaluated in Section~\ref{sec:Experiment}. On the other hand, traditional CS algorithms operate without ground-truth communities and thus demonstrate good generalization abilities to discover unseen communities. 
{\color{black}Therefore, a natural and promising idea is to develop a learning-based method that inherits the favorable properties of learning-based approaches, including flexible community structures, strong expressive capabilities and outstanding performance, and simultaneously combines the advantages of traditional CS algorithms, such as operating without ground-truth communities and demonstrating good generalization ability.
Hence, in this paper, we aim to design an efficient and effective learning-based approach for CS that runs without using ground-truth communities. The challenges are mainly two-folds: 1) Challenge I: effectively learning the community score for each query without using labels, and 2) Challenge II: adaptively identifying the community without label-based optimal threshold.}

One promising direction for community score learning without using labels is to use some unsupervised frameworks. However, the objective of existing unsupervised frameworks is incompatible with the task of community score learning, and existing works of CS cannot be easily extended to support unsupervised community score learning. 
Moreover, most of the existing unsupervised frameworks utilize message-passing-based Graph Neural Networks (GNNs) as the backbone. This choice inherently introduces challenges of the over-smoothing problem~\cite{chen2020measuring} and over-squashing problem~\cite{alon2021on} with the increment of model depth, consequently limiting their potential capability for graph representation learning as highlighted in~\cite{chen2022nagphormer}.
Therefore, designing a community score learning method that operates without using labels presents a considerable challenge.

One direct approach for community identification without using labels is to assign a fixed hyper-parameter as the threshold directly. However, as demonstrated in the evaluation conducted in ~\cite{jiang2022query}, utilizing different fixed thresholds can result in a maximum decrease of $\sim$40\% in the accuracy measured by the F1-score. Furthermore, the optimal threshold may vary across different datasets and different similarity metrics. Another promising approach is to select nodes with the top-\textit{K} highest community scores. However, the size of real-world communities can differ significantly within one graph and across different graphs. It is hard to utilize one fixed size that suits all the communities. Therefore, it is challenging to design a community identification method that runs without using labels.

\vspace{1mm}
\noindent \textbf{\color{black}{Our approaches.}} 
Driven by the aforementioned challenges, we propose a new pre-trained graph  \textbf{\underline{Trans}}former based community search framework that uses \textbf{\underline{Zero}} label (i.e., unsupervised), termed \framework. The overall illustration is in Figure ~\ref{fig:frameworks_comparison}(c). \framework~also contains two phases, i.e., the offline pre-training phase and the online search phase. The offline pre-training phase pre-trains the community search graph transformer (\transformername) designed specifically for CS, and the online search phase contains two key components, i.e., community score computation module and community identification with expected score gain (\searchname) solver module.

{\color{black}To address Challenge I, we introduce a two-step methodology that incorporates both the offline pre-training phase and the online community score computation module to obtain the community score without using labels. }
First of all, we pre-train the \transformername~in the offline pre-training phase. 
Subsequently, we calculate the community score in the community score computation module by measuring the similarity between the representation of the query and the representation of each node within the graph where the representation is inferred by the learned \transformername. 
{\color{black}Specifically, we propose \transformername~motivated by NAGphormer~\cite{chen2022nagphormer}, the existing state-of-the-art graph transformer. NAGphormer with other graph transformers have shown effectiveness for mitigating the over-smoothing and over-squashing problems~\cite{zhao2021gophormer, ying2021transformers, chen2022nagphormer}.}
To pre-train \transformername~without using labels, we introduce two self-supervised losses designed specifically for CS, i.e., personalization loss and link loss,  motivated by the inherent uniqueness of each node and graph topology, respectively. 
It combines the contrast-based self-supervised learning (i.e., personalization loss)~\cite{jiao2020sub, zhang2021motif} and the generation-based self-supervised learning (i.e., link loss)~\cite{liu2022graph, wu2021self} to achieve better performance. 

% \noindent \textit{(2) Maximum expected score gain solver module}. 
{\color{black}To solve Challenge II, we introduce a new function named expected score gain and formulate the problem of community identification as the problem of community \underline{I}dentification with \underline{E}xpected \underline{S}core \underline{G}ain (\searchname). }
Specifically, the community score reflects the likelihood of a node being included in the community, and the objective is to identify a community where nodes exhibit high scores. QD-GNN and COCLEP use a label-based threshold to define high scores. To eliminate reliance on labels, we define expected score gain. Community scores that maximize the expected score gain are considered high. Motivated by modularity~\cite{kim2022dmcs, schlosser2004modularity} which is a classical metric for community cohesiveness, the expected score gain calculates the sum of node scores within the community minus the sum of expected scores if nodes are chosen randomly. A higher expected score gain value indicates a potentially better community. 
Based on this new function and inspired by the cohesive nature of communities and the query-dependent property of CS, \searchname~ aims to find a connected subgraph that includes the query while having the maximum expected score gain value.
Furthermore, we prove that the \searchname~ is NP-hard and APX-hard, indicating that it cannot be solved in polynomial time and is inapproximable within any constant factor in polynomial time. Therefore, we design two heuristic algorithms, i.e., \localsearch~and \globalsearch~to effectively and efficiently identify promising communities without using labels.
% This formulation enables us to identify communities without relying on thresholds, and thus, without requiring labels.

% that aims to generalize to predict unseen communities

\vspace{1mm}
\noindent \textbf{Contributions.}
The main contributions are as follows:
\begin{itemize}[leftmargin=10pt, topsep=1pt]
\item{} We propose a new learning-based CS framework \framework~ that runs without using ground-truth communities. It contains the offline pre-training phase and the online search phase.
\item{} In the offline pre-training phase, we design an efficient graph transformer \transformername~for CS. Two self-supervised losses including the personalization loss and link loss are utilized to pre-train \transformername~without using labels.
{\color{black}\item{} In the online search phase, the score computation module first obtains the community score by measuring the similarity of the learned representation. Based on the new proposed expected score gain function, we model community identification as ~\searchname, and propose two efficient and effective algorithms, i.e., \localsearch~and \globalsearch, to find promising communities. }

\item{} Experiments across 10 public datasets highlight the superiority of \framework~regarding both accuracy and efficiency. 
Under the hybrid training setting, \framework~that does not use labels even outperforms COCLEP and QD-GNN which rely on labels with an average F1-score improvement of 10.01\% and 5.91\%, respectively. 
Regarding offline training efficiency, \framework~achieves an average speedup of 122.39$\times$ and 118.22$\times$ compared to COCLEP and QD-GNN, respectively. 
Regarding online search efficiency,  \framework~achieves an average speedup of 10.02$\times$ and 26.77$\times$ compared to COCLEP and QD-GNN, respectively.

\end{itemize}

\vspace{-4mm}
\section{Preliminaries}
\label{sec:Preliminary}

\begin{table}[t]
\centering %\small %\scriptsize
\caption{Symbols and Descriptions}
\vspace{-0.4cm}
\label{tab:symbol}
\begin{tabular}{|p{1.8cm}|p{6.0cm}|}
% {|p{3.2cm}|p{4.7cm}|} {|c|c|}
\hline
\cellcolor{lightgray}\textbf{Notation} & \cellcolor{lightgray}\textbf{Description} \\ \hline
$G(V, E)$ & an undirected graph\\ \hline
$X, A$ & feature matrix and adjacency matrix \\ \hline
$q=V_q$ & the query with node set $V_q$\\ \hline
$C_q, \tilde{C}_q$ & the ground-truth/predicted  community of $q$\\ \hline
$f^{\theta}(\cdot)$ & neural network model with parameters $\theta$ \\ \hline
$S\in \mathbb{R}^{|V|}$ & community score vector \\ \hline
$V_1 \backslash V_2$ & nodes in $V_1$ but not $V_2$ \\ \hline
% $ESG(S, C, G)$ & the expected score gain of $C$ given $S$ and $G$. \\ \hline
% $N^{(k)} (V_s)$ & the $k$-hop neighbor set of node set $V_s$ \\ \hline

\end{tabular}
\vspace{-5mm}
\end{table}

In this section, we give relevant preliminaries and the introduction of state-of-the-art models for CS. The frequently used symbols are summarized in Table~\ref{tab:symbol}.
\vspace{-3mm}
\subsection{Problem Statement}
% The $k$-hop neighbor of a node set $V_s$ is denoted as $N^{(k)} (V_s)$. 
We follow the typical setting of the general community search problem and focus on the undirected graph $G(V, E)$  where $V$ is the node set and $E\subseteq V\times V$ is the edge set. We use $\left|V\right|$ and $\left|E\right|$ to denote the cardinality of $V$ and $E$, respectively. The feature matrix is denoted as $X \in \mathbb{R}^{|V|\times d}$ where $d$ is the dimension of the feature. $A \in \mathbb{R}^{|V|\times |V|}$ is the adjacency matrix where  $A_{ij} = 1$ indicates the link between node $v_i$ and node $v_j$.  
$S\in \mathbb{R}^{|V|}$ is used to denote the community score vector.
We use $q$ and $V_q$ interchangeably to denote the query node set. $C_q$ and $\tilde{C}_q$ are utilized to denote the ground-truth community and the predicted community {w.r.t.} $q$. Next, we give the formal definition of community search.

\begin{definition}{(Community Search~\cite{li2023coclep, jiang2022query}).} 
Given a data graph $G(V, E)$ and query $q$, the task of Community Search (CS) aims to identify a query-dependent connected subgraph (i.e., community) $C_q$ where nodes in the found community are densely intra-connected.

\end{definition}

\vspace{-4mm}
\subsection{State-of-the-art}

The state-of-the-art learning-based CS models employ a framework comprising the offline training phase and the online search phase. 
% In this part, we discuss the details of these two stages.

\vspace{1mm}
\noindent \textbf{Offline Training}. QD-GNN models the problem of CS as binary node classification. 
Specifically,
given a data graph $G$ and a training dataset $\mathcal{D}_{train}=\{q_i, C_{q_i}\}_{i=1}^{|\mathcal{D}_{train}|}$ containing a set of queries and the corresponding ground-truth communities, QD-GNN proposes a neural network model, denoted as $\mathcal{M}$, which {\color{black}takes the query, adjacency matrix and features} as inputs and outputs a community score vector $S_q \in \mathbb{R}^{|V|}$. This vector indicates the membership likelihood of each node in the predicted community.
{\color{black}
\begin{equation*}
    S_q = \mathcal{M}(V_q, A, X)
\end{equation*}
}
And then, the Binary Cross Entropy (BCE) function is used to measure the BCE between $S_q$ and the ground-truth vector $Y_q$. $Y_{q,j} = 1$ if and only if $v_j \in C_q$. Here, $Y_{q,j}$ is the $j$-th bit of $Y_{q}$.

\begin{equation*}
    \mathcal{L} = \sum_{q\in \mathcal{D}_{train}} \frac{1}{|V|} \sum_{i=1}^{|V|}-\left(Y_{q,i} log S_{q,i}+\left(1-Y_{q,i}\right) log \left(1-S_{q,i}\right)\right)
\end{equation*}

The parameters of the model are updated by gradient descent to minimize the loss between the predicted community score and the ground-truth vector.
It is worth noting that the loss function employed in COCLEP differs from the aforementioned loss. COCLEP employs contrastive learning and focuses on enhancing the prediction performance for the selected positive candidates. However, both QD-GNN and COCLEP employ ground-truth information within their respective loss functions for model training.

Following the model training, both QD-GNN and COCLEP determine the optimal parameters, particularly the community score threshold, by evaluating the validation set. Note that, the validation set also contains the ground-truth information.

\vspace{1mm}
\noindent \textbf{Online search}. The learned model and the selected optimal threshold from the offline training phase are utilized in the online search phase. Specifically, it first calculates the community score by the model inference.
To identify the final community, QD-GNN proposes a Constrained Breadth-First Search algorithm in ~\cite{jiang2022query} which requires the label-based threshold as input. It expands outward and selects neighbors with a community score larger than the threshold.

\section{Offline Pre-training Phase}
\label{sec:Pretrain}

\noindent\textbf{Motivation.} 
Given a graph with feature matrix $X$ and adjacency matrix $A$, the objective of pre-training for CS is to learn a generic encoder that can encode the community information and the graph topology into the latent space. As we focus on the unsupervised CS, we resort to self-supervised learning which is an important category of unsupervised learning. Specifically, we consider both the generation-based self-supervised learning and the contrast-based self-supervised learning to achieve a better performance~\cite{qi2023recon}.

\vspace{-3mm}
\subsection{Overview}

\begin{figure}
  \centering
  \includegraphics[width=0.85\linewidth]{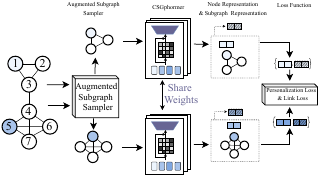}
  \vspace{-4mm}
  \caption{Illustration of the offline pre-training phase}
  \label{fig:pre_train_illustration}
\vspace{-6mm}
\end{figure}

The overall architecture of the offline pre-training phase is illustrated in Figure ~\ref{fig:pre_train_illustration}. Given a data graph, an augmented subgraph sampler is applied to generate the corresponding community-level subgraph.
It aims to generate new data with maximum consistent features from different views w.r.t. a node in the graph as positive samples and is used for contrastive training. 
Then, the augmented subgraph is sent into a graph encoder (\transformername~in our proposed \framework) to extract the latent features that encode the community information and graph topology. The graph encoder outputs both the node-level representation and community-level representation. The learned representations are used for loss computation which includes the personalization loss and the link loss. The obtained loss is back-propagated to update the parameters in \transformername.

% =============================================================================
\vspace{-2mm}
\subsection{Augmented Subgraph Sampler}
\noindent\textbf{Motivation.} The contrast-based self-supervised learning is based on the augmented subgraphs. COCLEP, an existing state-of-the-art semi-supervised CS model, constructs the augmented subgraphs by incorporating the ground-truth positive samples and their $K$-hop neighbors, where $K$ is a fixed value. However, it is intuitive that various central nodes should have different neighborhood distances, necessitating a personalized selection of different hops for different nodes. To address this, we employ conductance~\cite{andersen2006local, yang2012defining}, a well-established measure of community cohesiveness, to enhance the choice of hops. The definition of conductance is as follows:
\begin{definition}{(Conductance~\cite{andersen2006local, yang2012defining}).}
Given a graph $G(V, E)$ and a community $C$, the conductance of $C$ is defined as: 
\begin{equation}
    \label{equ:density_modularity}
    \Phi(G, C)=\frac{|e(C, \overline C)|}{min(d_C, d_{\overline C})}
\end{equation}
where $\overline C=V\backslash C$ is complement of $C$. $e(C, \overline C)$ is the edges between nodes in $C$ and nodes in $\overline C$.  $d_C$ is the sum of degrees of the nodes in $C$. 
\end{definition}

Conductance measures the fraction of the total edge volume that points outside the community. A smaller conductance means a higher ratio of information can be used for pre-training.
Based on conductance, we choose the subgraph induced by the $K$-hop neighbors of the query nodes that has the lowest conductance value as the augmented subgraph. It's important to note that this approach allows us to obtain the augmented subgraph adaptively, without requiring a pre-set value for $K$. To strike a balance between search space and personalization, we set the upper limit for $K$ as 5 as suggested by the experimental results.

{\color{black}
\begin{example}
    Given the data graph as in Figure~1 and node 1, its 1-hop induced subgraph has nodes 1, 2 and 3. Thus, the conductance of the 1-hop induced subgraph is $\frac{1}{7}=0.143$. Similarly, the conductance of the 2-hop induced subgraph is $\frac{3}{9}=0.333$. Thus, 1-hop induced subgraph is selected as the augmented subgraph.
\end{example}
}
% =============================================================================
\vspace{-2mm}
\subsection{\transformername~Architecture}

\noindent\textbf{Motivation.} The graph encoder used in the pre-training phase inputs the augmented subgraph and outputs both the node-level and the community-level representations. A direct way is to use GNNs to learn the node-level representation and use a graph pooling operator to aggregate the node-level representations into community-level representations. However, GNNs possess inherent limitations such as over-smoothing~\cite{chen2020measuring} and over-squashing~\cite{alon2021on} issues, which hinder the full potential of GNNs for representation learning. On the other hand, transformers have been recently introduced for graph analytics due to their effectiveness in addressing over-smoothing and over-squashing issues~\cite{chen2022nagphormer}, resulting in a strong representation learning capacity. Many graph transformer models are proposed such as Graphormer~\cite{ying2021transformers} and Gophormer~\cite{zhao2021gophormer}. 

{\color{black}Here, we follow the state-of-the-art graph transformer NAGphormer~\cite{chen2022nagphormer} and propose \transformername. 
NAGphormer is the state-of-the-art graph transformer with high efficiency. NAGphormer treats each hop as one token in a sequence and uses a transformer to model the correlation among different hops and learn the node representation.}
\transformername~has three key distinctions from NAGphormer. Firstly, NAGphormer primarily targets supervised learning, while \transformername~focuses on unsupervised learning. Secondly, unlike the fixed value of $K$ used for $K$-hop neighbors in NAGphormer, we employ the conductance to dynamically determine the value of $K$. Thirdly, \transformername~outputs both the node-level and the community-level representations while NAGphormer just outputs the node-level representation.

\begin{figure}
  \centering
  \includegraphics[width=0.91\linewidth]{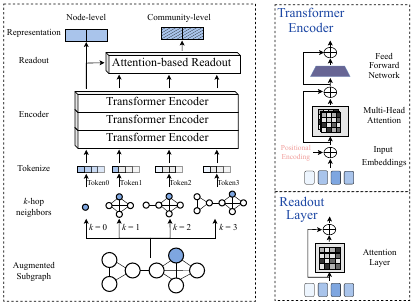}
  \vspace{-5mm}
  \caption{Architecture of \transformername}
  \label{fig:CSGphormer}
\vspace{-6mm}
\end{figure}

The architecture of \transformername~is illustrated in Figure ~\ref{fig:CSGphormer}, and the forward propagation of  \transformername~ is summarized in Algorithm~\ref{algo:CSGphormer}. Specifically, we propagate the feature matrix $X$ from 1 to $K$ times to obtain the token sequence $\mathcal{X}=\{^0X,~^1X, \cdots, ~^KX\}$. Here, $^0X = X \in \mathbb{R}^{n\times d}$ is the original feature matrix. $^kX \in \mathbb{R}^{n\times d}$ is the $k$-hop neighborhood matrix and is computed by $^kX = \hat{A}^k X$. Here,  $\hat{A} = D^{-\frac{1}{2}}AD^{-\frac{1}{2}}$ is the normalized adjacency matrix. $D$ is the degree matrix of $A$ where $D(i,i) = \sum_{j=1}^{n}A(i, j)$. $\mathcal{X}_v=\{^0x_v, \cdots,  ~^{K}x_v \}$ is the aggregated neighborhood sequence of node $v$.

Then, the obtained sequence matrix $\mathcal{X}_v \in \mathbb{R}^{(K+1)\times d}$ is sent into a learnable linear projection $W \in \mathbb{R}^{d\times d_m^{(0)}}$: $H^{(0)}_v = \mathcal{X}_v W $,
where $H^{(0)}_v \in \mathbb{R}^{(K+1)\times d_m^{(0)}}$. Next, $H^{(0)}_v$ is sent for $L$ layers of the transformer encoder. The transformer encoder contains three important sub-structures, i.e. Positional Encoding, Multi-Head Attention and Feed Forward Network. Specifically, given an input embedding $H^{(l)}_v $ in layer $l$, the position encoding of the sequence $P$ is added to the input embedding first $H^{(l)}_v = H^{(l)}_v+P$, as in ~\cite{chen2022nagphormer}.

After that, $H^{(l)}_v$ is sent to the multi-head attention layer (MHA): 
\begin{equation}
\begin{aligned}
    \textrm{MHA}(H^{(l)}_v) = \textrm{Concat}(\textrm{head}_1, \cdots,\textrm{head}_h)W^o \\
    \textrm{where}~\textrm{head}_i = \textrm{Attention}(H^{(l)}_vW^q_i, H^{(l)}_vW^k_i, H^{(l)}_vW^v_i) \\
    \textrm{and} ~ \textrm{Attention}(Q, K, V) = \textrm{softmax}(\frac{QK^T}{\sqrt{d_m^{(l+1)}}})V
\end{aligned}
\end{equation}
Here, $W^q_i \in \mathbb{R}^{d_m^{(l)}\times d_m^{(l+1)}}$, $W^k_i \in \mathbb{R}^{d_m^{(l)}\times d_m^{(l+1)}}$, $W^v_i \in \mathbb{R}^{d_m^{(l)}\times d_m^{(l+1)}}$, $W^o \in \mathbb{R}^{hd_m^{(l+1)}\times d_m^{(l+1)}}$ are all learnable weights to project $H^{(l)}_v$ into different matrices. 

The output of the multi-head self-attention layer is added to the original input embedding followed by a layer normalization (LN)~\cite{ba2016layer}. A position-wise feed-forward network (FFN) is applied to each position separately and identically. The FFN consists of two linear layers with a GELU~\cite{hendrycks2016gaussian} non-linearity:

\begin{equation}
\begin{aligned}
H^{(l+1)}_v = \textrm{MHA}(\textrm{LN}(H^{(l)}_v))+H^{(l)}_v \\
H^{(l+1)}_v = \textrm{FFN}(\textrm{LN}(H^{(l+1)}_v ))+H^{(l+1)}_v
\end{aligned}
\end{equation}

After the $L$ layers of the transformer encoder, we can obtain the latent representation $H^{(L)}_v\in \mathbb{R}^{(K+1)\times d_m^{(L)}}$ of node $v$ which contain the center node token $Z^{node}_v = {^0}H^{(L)}_v \in \mathbb{R}^{d_m^{(L)}}$ and the latent representation of the neighborhood tokens $\{^1H^{(L)}_v, \cdots,~^KH^{(L)}_v\}$. Then, an attention-based readout function is utilized to weight the neighborhood tokens to obtain the community-level representation:

\begin{equation}
\begin{aligned}
% \begin{gather*}
&\alpha_k = \frac{\textrm{exp}((^0H^{(L)}_v|| ^kH^{(L)}_v)W_a)}{\sum_{i=1}^K\textrm{exp}((^0H^{(L)}_v|| ^iH^{(L)}_v)W_a)}\\
&Z^{com}_v = \sum_{k=1}^K \alpha_k~^kH^{(L)}_v.
% \end{gather*}
\end{aligned}
\end{equation}
Here, $||$ is the concatenation operator, and $W_a \in \mathbb{R}^{ 2d_m^{(L)} \times 1}$ is the learnable weight matrix.

% =============================================================================

\begin{algorithm}[t]
\captionsetup{singlelinecheck=false} % 防止标题换行
\captionsetup{margin={0pt,5em}}
% \caption{ Constrained BFS for community identification}
\caption{\parbox{\linewidth}{Forward Propagation of \transformername.}}
\label{algo:CSGphormer}
\LinesNumbered
\DontPrintSemicolon
\KwIn{center node $v$, feature matrix $X$, adjacent matrix $A$, transformer layers $L$. }
\KwOut{The node representation $Z_v^{node}$ and community-level representation $Z_v^{com}$ .}

$\mathcal{X}_v \leftarrow \{^0x_v, ~^{1}{x}_v, \cdots,  ~^{K}x_v \} $ \;

$ H^{(0)}_v \leftarrow \mathcal{X}_v W $ \;

\tcp*[l]{L-layers transformer encoder.}

\For{$l=0,\cdots, L-1$}{
    $ P \leftarrow $ Position Encoding Construction \;
    $ H^{(l)}_v \leftarrow H^{(l)}_v + P$ \;
    $H^{(l+1)}_v = \textrm{MHA}(\textrm{LN}(H^{(l)}_v))+H^{(l)}_v$ \;
    $H^{(l+1)}_v = \textrm{FFN}(\textrm{LN}(H^{(l+1)}_v ))+H^{(l+1)}_v$ \;
    
}

\tcp*[l]{Readout layer.}

$Z_v^{node} \leftarrow {^{0}}H^{(L)}_v$; $Z_v^{com} \leftarrow \textrm{Zero Tensor}$

\For{$k=1,\cdots, K$}{
    $\alpha_k = \frac{\textrm{exp}((^0H^{(L)}_v|| ^kH^{(L)}_v)W_a^T)}{\sum_{i=1}^K\textrm{exp}((^0H^{(L)}_v|| ^iH^{(L)}_v)W_a^T)}$ \;
    $Z_v^{com} \leftarrow Z_v^{com} + \alpha_k~^kH^{(L)}_v $
}

\Return $Z_v^{node}, Z_v^{com}$\;
\end{algorithm}

\vspace{-3mm}
\subsection{Training Objectives}

\noindent\textbf{Motivation.} Intuitively, nodes are dependent on their communities to learn the representation and each node is unique in the graph. We consider the strong correlation between central nodes and their communities to design the contrast-based loss, i.e., personalization loss. Moreover, we design the generation-based loss, i.e., link loss, based on the idea that \textit{nodes that have a link should be close in the latent space and vice versa.} The generation-based loss would benefit the preservation of local graph topology, and the contrast-based loss would benefit the preservation of the global information and capture the long-distance relationship, as illustrated in ~\cite{qi2023recon}. 

We model the personalization loss by the margin triplet loss~\cite{schroff2015facenet} to bring the representation of a selected node and its corresponding community closer together and push away the representation of the selected node from the communities of other nodes. 
% The computation of the margin triplet loss is defined as:

\begin{equation}
    \mathcal{L}_p = \frac{1}{|V|^2}\sum_{v \in V}\sum_{u \in V} \left(-\textrm{max}\left(\sigma(Z^{node}_v Z^{com}_v)-\sigma(Z^{node}_v Z^{com}_u)+\epsilon, 0\right)\right)
\end{equation}
where $\epsilon$ is the margin value and $\sigma(\cdot)$ is the sigmoid function.

The link loss is formulated as follows to enhance the similarity of neighboring nodes while discriminating non-adjacent nodes.

\begin{equation} 
    \begin{aligned}
    \mathcal{L}_k = \frac{1}{|V|^2}\sum_{v \in V}\sum_{u \in V} -A(u,v)  (Z^{node}_u Z^{node}_v)\\
    +(1-A(u,v))(Z^{node}_u Z^{node}_v)
    \end{aligned}
\end{equation}

\transformername~takes the above two losses into account together. The overall loss function is defined as:

\begin{equation}
\label{equ:loss_func}
    \mathcal{L} = \mathcal{L}_p+\alpha\mathcal{L}_k
\end{equation}
where $\alpha \in [0,1]$ is the coefficient to balance the two losses. Note that both $\mathcal{L}_p$ and $\mathcal{L}_k$ do not contain label information.

With the loss defined in Equation~\ref{equ:loss_func}, the overall offline pre-training procedure is summarized in Algorithm~\ref{algo:training_procedure}. It first initializes the parameters in the optimizer and divides all nodes into several batches (lines 1 to 2). Next, it samples all the augmented subgraphs (lines 3 to 4). It trains the model batch by batch. In the training of each batch, it obtains the node-level representation and community-level representation of all nodes in the batch by propagating the \transformername~network (lines 6 to 7). The loss containing both the personalization loss and link loss is computed (lines 8 to 11) and is used to update the parameters in \transformername (line 12).

\begin{algorithm}[t]
\captionsetup{singlelinecheck=false} % 防止标题换行
\captionsetup{margin={0pt,5em}}
% \caption{ Constrained BFS for community identification}
\caption{\parbox{\linewidth}{Offline Pre-training Procedure (One Epoch)}}
\label{algo:training_procedure}
\LinesNumbered
\DontPrintSemicolon
\KwIn{The data graph $G$, batch size $n_{batch}$, layer number $L$, \transformername~ $f^{\theta}(\cdot)$, learning rate $\varphi$, coefficient $\alpha$. }
% \KwOut{The learned.}

Initialize optimizer $opt_{\theta}$ with learning rate $\varphi$\;
Separate $V$ into batches $\{V_b\}$ with the batch size $n_{batch}$\;

\For{each $v \in V$}{
     $G_{v} \leftarrow $ augmented subgraph sampler\;
}

\For{$\{V_b\} \in V$}{
     % \For{each $v \in V_b$ and $v \in V_b$ and $v \neq u$}{
     \For{each $v \in V_b$}{
     $Z_v^{node}, Z_v^{com} \leftarrow f^{\theta}(v, X(G_v), A(G_v), L)$
     }
     % $\mathcal{L} \leftarrow 0$ \;
    %  $\mathcal{L}_p = \frac{1}{|V_b|^2}\sum_{v \in V_b}\sum_{u \in V_b} \left(-\textrm{max}\left(\sigma(Z^{node}_v Z^{com}_v)+\sigma(Z^{node}_v Z^{com}_u)+\epsilon, 0\right)\right)$ \;
    %  $\mathcal{L}_k = \frac{1}{|V_b|^2}\sum_{v \in V}\sum_{u \in V} -A(u,v)  (Z^{node}_u Z^{node}_v)
    % +(1-A(u,v))(Z^{node}_u Z^{node}_v)$\;
     \For{each $u, v \in V_b$}{
     $\mathcal{L}_p = -\textrm{max}\left(\sigma(Z^{node}_v Z^{com}_v)-\sigma(Z^{node}_v Z^{com}_u)+\epsilon, 0\right)$\;
     $\mathcal{L}_k = -A(u,v)  (Z^{node}_u Z^{node}_v)+(1-A(u,v))(Z^{node}_u Z^{node}_v)$\;
     $\mathcal{L} += \mathcal{L}_p+\alpha\mathcal{L}_k$\;
     }
     Update $\theta$ by $opt_{\theta}$ with loss $\frac{\mathcal{L}}{\left|V_b\right|^2}$.\;
}

% \For{each $u \in N(v)$ and $S_{q,u} \geq \gamma$}{
%      $Q \leftarrow Q \cup u$; $\Tilde{C}_q \leftarrow \Tilde{C}_q \cup u$\;
% }

\end{algorithm}

\vspace{-2mm}
\section{Online Search Phase}
\label{sec:Search}
\vspace{-1mm}

\begin{algorithm}[t]
\captionsetup{singlelinecheck=false} 
\captionsetup{margin={0pt,5em}}
\caption{\parbox{\linewidth}{Community Score Computation}}
\label{algo:score_computation}
\LinesNumbered
\DontPrintSemicolon
\KwIn{The query $V_q$, graph $G$, pre-trained network $f^{\theta}(\cdot)$ .}
\KwOut{The community score $S$.}

Initialize $S\leftarrow \{s_v =0\ \textrm{for}\ v \in V\}$\;

\For{$\{v\} \in V$}{
     \For{$\{u\} \in V_q$}{
        $s_v \leftarrow s_v + \frac{\sum_{i=0}^{d_m^{(L)}}f^{\theta}_i(v)f^{\theta}_i(u)}{\sqrt{\sum_{i=0}^{d_m^{(L)}}f^{\theta}_i(v)f^{\theta}_i(v)}\times \sqrt{\sum_{i=0}^{d_m^{(L)}}f^{\theta}_i(u)f^{\theta}_i(u)}}$
}
${s_v}\leftarrow\frac{s_v}{|V_q|} $;
}
\Return $S$\;

\end{algorithm}

With the pre-trained \transformername~ in Section~\ref{sec:Pretrain}, we now introduce the details of the online search phase devised for unsupervised CS. We first introduce the community score computation module, and then we introduce the problem of identification with expected score gain (\searchname), followed by two search algorithms, i.e., \localsearch~and \globalsearch~to find promising communities.
% \subsection{Overview}
\vspace{-2mm}
\subsection{Community Score Computation}
\vspace{-1mm}
\noindent\textbf{Motivation.} After pre-training, the community-level information and the graph topology are encoded into the latent representation. Nodes with similar latent representations should have similar community-level information and should be close to each other in the original graph. 
Therefore, we can compute the community score by evaluating the similarity between the representation of the query and the representations of nodes within the graph.
A higher similarity suggests a greater likelihood of the node being part of the resulting community.

The overall community score computation algorithm is shown in Algorithm~\ref{algo:score_computation}. It inputs the query nodes, graph and the pre-trained graph transformer and outputs the community score {w.r.t.} the query. The community score is first initialized as all zeros (line 1). It then computes the pairwise similarity between the representation of each node in the query and the representation of each node in the graph (lines 2 to 5). We use the cosine similarity here. More other similarity functions are evaluated in the experiments of Section~\ref{sec:Experiment}. We ensure that the obtained score is adjusted to fall within the range of 0 to 1 by normalizing it with the cardinality of the query node set (line 5). At last, the community score is returned.

\vspace{-2mm}
\subsection{Identification with Expected Score Gain}
\vspace{-1mm}
\noindent\textbf{Motivation.} The community score quantifies the likelihood of a node being included in the community. An ideal community is one where all nodes exhibit high community scores w.r.t. the query nodes. 
In order to measure the degree to which the score is large, QD-GNN and COCLEP use the label-based threshold, and nodes having a community score larger than the label-based threshold are included in the resulting communities. 
{\color{black}Under the setting of unsupervised community search, as analyzed in Section~\ref{sec:Introduction}, a naive approach like a fixed threshold or a fixed number of nodes would potentially harm the accuracy of community identification. Therefore, we introduce the function of expected score gain (\textit{ESG}), and community scores that maximize the \textit{ESG} are considered high. The \textit{ESG} function, which centers around utilizing community scores, computes the sum of node scores within the community, subtracted by the sum of expected scores under random node selection. A higher \textit{ESG} suggests the potential for a superior community. This idea is inspired by the concept of modularity, a well-established metric of community cohesiveness.} The modularity measures the number of edges in the community minus the expected number of edges in the community if the edges are randomly distributed. The higher the modularity, the more cohesive the community~\cite{kim2022dmcs}. 

\begin{definition}{(Expected Score Gain).}
\label{definition:esg}
Given a graph $G(V, E)$, a community $C=(V_C, E_C)$ and the community score $S$, the expected score gain of $C$ is defined as: 
\begin{equation}
    \label{equ:density_modularity}
    \textrm{ESG}(S, C, G)=\frac{1}{|V_C|^\tau}(\sum_{v\in V_C}s_v-\frac{\sum_{u\in V}s_u}{|V|}|V_C|)
\end{equation}
where $\tau \in [0, 1]$ is a hyper-parameter to control the granularity of the subgraph, and a higher $\tau$ value leads to a more fine-grained subgraph.  
\end{definition}

The first term $\sum_{v\in V_C}s_v$ is the sum of the community score of nodes in the selected community. $\frac{\sum_{u\in V}s_u}{|V|}|V_C|$ is the expected community score where nodes randomly selected have an expected score the same as the average score of the graph. We set $\tau = 0.5$ as suggested by our experiments in Section~\ref{sec:Experiment}. 
{\color{black}\begin{example}
    Given the data graph as in Figure~1 with the community containing nodes 4 5, 6 and 7, and supposing the community scores are 0.1, 0.2, 0.4, 0.7, 0.9, 0.6, 0.8 for nodes 1 to 7 respectively, the expected score gain can be calculated as $\frac{1}{4^{0.5}}(3.0-\frac{3.7}{7}\times4)=0.443$.
\end{example}
}
Besides the preference for nodes in the resulting community to exhibit high community scores, we aim to search for a cohesive subgraph, and connected components are more cohesive than unconnected subgraphs. Furthermore, the identified community is query-dependent and should therefore include the query nodes. Based on these considerations, we formally define the problem of identification with expected score gain (\searchname).
\begin{definition}{(Identification with Expected Score Gain).}
Given a graph $G(V, E)$, the query $V_q$, the community score $S$ and a profit function $\textrm{ESG}(\cdot)$, \searchname~aims to select a community $C$ of $G$, such that:\begin{enumerate}[topsep=1pt]
\item {\color{black}$V_C$ contains nodes in $V_q$, and $C$ is connected;} 
\item $\textrm{ESG}(S, C, G)$ is maximized among all feasible choices for $C$.
\end{enumerate}
\end{definition}

We give the hardness of \searchname~in Lemma~\ref{lemma:np_hard}, and the proof can be found in Section~\ref{sec:Analysis}.

\begin{lemma}
\label{lemma:np_hard}
The problem of \searchname~ is NP-hard.
\end{lemma}

Then, we prove that there is no polynomial time approximation scheme (PTAS) for \searchname, and thus it is APX-hard unless P=NP. The proof is also in Section~\ref{sec:Analysis}.

\begin{lemma}
\label{lemma:apx_hard}
For any $\epsilon \textgreater 0$, the \searchname~problem cannot be approximated in polynomial time within a ratio of {\color{black}$(1-\epsilon)\ln (|V|)$}, unless P=NP. 
\end{lemma}

\vspace{-4mm}
\subsection{\color{black}Heuristic Algorithms}

\begin{algorithm}[t]
\captionsetup{singlelinecheck=false} 
\captionsetup{margin={0pt,5em}}
\caption{\parbox{\linewidth}{\localsearch~Algirithm}}
\label{algo:localsearch}
\LinesNumbered
\DontPrintSemicolon
\KwIn{The community score $S$, graph $G$ and query $V_q$.}
\KwOut{The identified community $\Tilde{C}_q$.}

% $C\leftarrow$ Compute the Steiner tree of $V_q$\;
$\Tilde{C}_q, Q\leftarrow V_q$; $max\_esg\leftarrow -inf$\;

% $S_{temp}=\{s_i|v_i\in \Tilde{C}_q \cup \{u \}\}$
\While{$\left|Q \right| \textless \left|V \right| $}{
    $u \leftarrow \textrm{argmax}_{v \in \zeta \overline Q} s_v$ \;
    $Q = Q \cup u$;  \;
    \If{$\textrm{ESG}(S, \Tilde{C}_q \cup \{u \}, G) > max\_esg$}{
         $max\_esg \leftarrow \textrm{ESG}(S, \Tilde{C}_q \cup \{u \}, G)$\;
         {\color{black}{$\Tilde{C}_q=\Tilde{C}_q \cup \{u \}$}}\;
    }
    \Else{
        Terminate\;
    }
}

\Return $\Tilde{C}_q$\;

\end{algorithm}

{\color{black}As \searchname~ is NP-hard and APX-hard which means it cannot be solved in polynomial time and is inapproximable within any constant factor in polynomial time, heuristic algorithms are proposed to effectively and efficiently find promising communities without using labels. A direct approach (\localsearch) starts from the query nodes and greedily incorporates the node with the highest community score that is in the neighborhood of the selected intermediate subgraph. The subgraph with the largest expected score gain encountered during the search process is returned. In this way, we do not need to use the label-based threshold, and thus this algorithm does not require the labels for community identification. }

The overall algorithm of \localsearch~ is summarized in Algorithm~\ref{algo:localsearch}. It inputs the community score, graph and query nodes, and outputs the identified community. We use $Q$ to store the nodes that have been traversed and designate the maximum expected score gain value as negative infinity during initialization (line 1). The algorithm terminates until all the nodes have been traversed or early stops when there are no promising candidates (lines 2 to 9). In each loop, we first select the node with the highest community score that has not been traversed and is located at the boundary of the node set that has been traversed (line 3). $\zeta \overline Q = \{v \in \overline Q | \exists i \in N^{(1)}(v) \cap Q \}$ is the boundary of $\overline Q$ and  $\overline Q=V\backslash Q$. If the selected nodes can increase the expected score gain of the previous intermediate subgraph, we incorporate it into the community. Otherwise, the algorithm early stops (lines 5 to 9). At last, the community returns.

\vspace{1mm}
\noindent\textbf{Motivations for \globalsearch.} 
{\color{black}
The motivations behind \globalsearch~can be outlined in three aspects. 
Firstly, by incorporating link loss, the learned representation effectively preserves graph topology information. Nodes with high similarities to the query nodes exhibit high community scores and are likely connected to the query nodes. This suggests that optimizing the expected score gain first can also provide a favorable priority for connected nodes. Secondly, the time complexity of \localsearch~is $O(|V|^2\log(|V|))$, as detailed in Section~\ref{sec:Analysis}. This quadratic logarithmic time complexity presents challenges when applying \localsearch~to large graphs. Thirdly, as detailed in Lemma~\ref{lemma:global_decrase}, the expected score gain of the first $p$ nodes in the queue sorted by community score initially increases and then decreases with the increase of $p$.}

{\color{black}
\begin{lemma}
\label{lemma:global_decrase}
Given sorted scores $\hat{S}$ from large to small, size $ p\textgreater0, C_{p} = \{v_i| \hat{s}_i \geq \hat{s}_{p}\}, S_{p} = \{\hat{s}_i| i \leq {p}\}$,  assuming $\sum_{\hat{s}_i\in S_{p}}\hat{s}_i=\mu|S_{p}|^{\sigma\tau}$ where $\mu, \sigma$ are hyperparmeters and $\mu, \sigma \textgreater 0$ and, $\sigma\tau \textless 1$, $ESG(\hat{S}, C_p, G)$ first increases and then decreases as $p$ increases.
\end{lemma}

As $S$ is learned, there lacks a functional expression for $S$. Since $S_p$ gets the first $p$ scores from a sorted queue $\hat{S}$, we assume that the sum of ${S}_p$ exhibits a decreasing growth rate as size increases, i.e. $\sum_{\hat{s}_i\in S_{p}}\hat{s}_i=\mu|S_{p}|^{\sigma\tau}$. Lemma~\ref{lemma:global_decrase} provides theoretical support for the binary search optimization, and the proof is in Section~\ref{sec:Analysis}.
}

{\color{black}Given these motivations and the theoretical foundation, we introduce \globalsearch~in Algorithm~\ref{algo:globalsearch}, which prioritizes candidates that enhance the \textit{ESG} from a global perspective. }
It inputs the learned community score, graph and query nodes, and outputs the found community. The community is initialed as the query nodes, and we designate the search start point $t_s$ as $0$ and the search end point $t_e$ as the maximum number of nodes during initialization (line 1). The algorithm sorts all the community scores (line 2). It loops until the start point equals the end point (lines 3 to 9). In each loop, it selects the candidate community $C_{mid}$ from 0 to $\frac{t_s+t_e}{2}$ and selects the candidate community $C_{left}$ from 0 to $\frac{t_s+t_e}{2}-1$ which lies at the left of $C_{mid}$ (lines 4 and 5). If the \textit{ESG} of $C_{mid}$ is larger than that of $C_{left}$ which means there may be promising candidates in the index range of $[\frac{t_s+t_e}{2},t_e]$, we set the new start point as $\frac{t_s+t_e}{2}$ (lines 6 and 7). Otherwise, the end point is set as $\frac{t_s+t_e}{2}$ (lines 8 and 9). At last, the identified community returns (line 10).

{\color{black}
While the community found by \globalsearch~may not always be guaranteed to be connected, it prioritizes connected nodes since link loss is used for pre-training, as shown in our earlier motivation. Additionally, \globalsearch~has a time complexity of $O(2 \times |V|\log(|V|))$, as analyzed in Section~\ref{sec:Analysis}, making it well-suited for large datasets.
}

\begin{algorithm}[t]
\captionsetup{singlelinecheck=false} 
\captionsetup{margin={0pt,5em}}
\caption{\parbox{\linewidth}{\globalsearch~ Algorithm}}
\label{algo:globalsearch}
\LinesNumbered
\DontPrintSemicolon
\KwIn{The community score $S$, graph $G$ and query $V_q$.}
\KwOut{The identified community $\Tilde{C}_q$.}

$\Tilde{C}_q\leftarrow V_q$;$t_s= 0$; $ t_e = \left|{S} \right|$\;
$\hat{S} \leftarrow$ sort $S$ from large to small\;
% ; 
%     ${S}_{mid} = \{\hat{s}_i| i \leq \frac{t_s+t_e}{2}\} $
% ; 
%     ${S}_{left} = \{\hat{s}_i| i \leq \frac{t_s+t_e}{2}-1\} $
\While{$t_s \textless t_e $}{
    $C_{mid} = \{v_i| \hat{s}_i \geq \hat{s}_{\frac{t_s+t_e}{2}}\}$\;  
    $C_{left} = \{v_i| \hat{s}_i \geq \hat{s}_{\frac{t_s+t_e}{2}-1}\}$\;  
    % $C_{right} = \{v_i| \hat{s}_i \geq \hat{s}_{\frac{t_s+t_e}{2}+1}\}$; 
    % ${S}_{right} = \{\hat{s}_i| i \leq \frac{t_s+t_e}{2}+1\} $\; 
    \If{$\textrm{ESG}(\hat{S}, C_{mid}, G) > \textrm{ESG}(\hat{S}, C_{left}, G)$}{
         $t_s \leftarrow \frac{t_s+t_e}{2}$\;
    }
    \Else{
         $t_e \leftarrow \frac{t_s+t_e}{2}$\;
    }
}
\Return $\Tilde{C}_q = \Tilde{C}_q \cup \{v_i| \hat{s}_i \geq \hat{s}_{t_e}\}$\;
\end{algorithm}

\begin{lemma}
\label{lemma:global}
\globalsearch~runs at most $log_2(|V|)$ iterations.
\end{lemma}

The proof of Lemma~\ref{lemma:global} is immediate as \globalsearch~reduces the search space by half each iteration.

{\color{black}
\begin{example}
    Given query node 5, and data graph with community scores same as in Example~2, \localsearch~ first selects node 7 as it has the highest community score of 0.8 and is the neighbor of node 5.
\end{example}

\begin{example}
    Given query node 5, and data graph with community scores same as in Example~2, \globalsearch~first sorts nodes according to community scores. In the first iteration, $C_{mid}$ are nodes $\{5, 7, 4, 6\}$ with \textit{ESG} 0.443 and $C_{left}$ are nodes $\{5, 7, 4\}$ with \textit{ESG} 0.470. $0.470>0.443$. Thus, the end point is halved.
\end{example}
}

\vspace{-2mm}
\section{Analysis}
\label{sec:Analysis}
\vspace{-1mm}

% {\color{blue}\noindent \textbf{Future works}: \framework~ is designed for general CS without using labels. However, \framework~can be easily extended to support other types of graph data and query inputs. With specific pre-training methods, \framework~can be extended for CS of hypergraph~\cite{lee2023k}, multiplex graph~\cite{behrouz2022cs}, bipartite graph~\cite{wang2021efficient} and temporal graph~\cite{hashemi2023cs}. With specific score computation methods, \framework~can be extended for CS for attributed community search~\cite{jiang2022query}. Moreover, online search algorithms can be further improved with specific settings and related theoretical works in~\cite{jin2021unconstrained}.}
% Here, we prove that the problem of \searchname~is NP-hard.
\subsection{Theoretical Analysis}
\vspace{-1mm}
\noindent \textbf{Proof of Lemma 1.} We reduce the problem of \searchname~ from the set cover problem which is a well-known NP-hard problem~\cite{arora1998approximability}. The following gives the formal definition of the set cover problem.
%of $|M|$ elements
\begin{definition}[Set Cover Problem]
{\color{black}
Given a finite set $M = \{m_1, m_2, \ldots, m_{|M|}\} $ and a collection $\mathcal{N}=\{N_1, N_2, \ldots, N_{|\mathcal{N}|}\}$ of subsets of $M$, the Set Cover Problem aims to find a minimum-size subcollection $\mathcal{N}_{opt}$ such that the union of all sets in $\mathcal{N}_{opt}$ covers all elements in $M$, i.e., $\bigcup_{N_i \in \mathcal{N}_{opt}} N_i = M$.
}
\end{definition}

\begin{figure}
  \centering
  \includegraphics[width=0.90\linewidth]{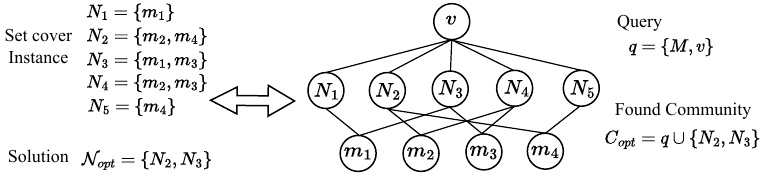}
  \vspace{-4mm}
  \caption{Graph construction for the set cover problem}
  \label{fig:nophard}
\vspace{-5mm}
\end{figure}

% We then construct a graph as follows. For each element in $m_i \in M$, we create two nodes, i.e., $m_i$ and $z_i$. One edge is added between  $m_i$ and $z_i$. We create a node for each set in $N$, and one edge is added between  $m_i$ and $n_i$ if $m_i \in n_i$. Another node $v$ is added with edges between $v$ and each set in $N$. Figure~\ref{fig:nophard} gives an example of graph construction. The set cover instance is $n_1 = \{m_1\}, n_2 = \{m_2, m_4\}, n_3 = \{m_1, m_3\}, n_4 = \{m_2, m_3\}, n_5 = \{m_4\}$.
We then follow~\cite{wu2015robust} to construct a graph. We create one node $m_i$ for each element in $m_i \in M$ and a node $N_i$ for each set in $N_i \in \mathcal{N}$. One edge is added between  $m_i$ and $N_i$ if $m_i \in N_i$. Another node $v$ is added with edges between $v$ and each set in $\mathcal{N}$. Figure~\ref{fig:nophard} presents an example of graph construction. The set cover instance is $N_1 = \{m_1\}, N_2 = \{m_2, m_4\}, N_3 = \{m_1, m_3\}, N_4 = \{m_2, m_3\}, N_5 = \{m_4\}$.

We use $w(C)=\sum_{v\in C}s_v$ to denote the sum of scores in $C$ and set $\tau = 1$ in $\textit{ESG}(S, C, G)$. Therefore, we have $\textit{ESG}(S, C, G) = \frac{w(C)}{|V_C|}-\frac{w(G)}{|V|}$. We omit the second term as it is a fixed value for communities in one graph. We use $g(\cdot)$ to denote the simplified function, i.e., $g(S, C, G) = \frac{w(C)}{|V_C|}$ and assume the community score of nodes in $M$ and $\{v\}$ is $\frac{1}{\left| M \right|+1}$, and the community score of nodes in $\mathcal{N}$ is $\frac{1}{\left|M\right|\left|\mathcal{N}\right|}$. 

% We set the query nodes as $q = Z\cup M\cup\{v\} $. To make the return community connected, we need to select some nodes in $N$. Let $N_*\subseteq N$ be a feasible solution for the set cover problem given the element collection $M$ and the set collection $N$. 
% Let $C = q\cup N_* = Z\cup M\cup\{v\} \cup N_*$. Then, C is connected and has the expected weight gain $g(S, q\cup N_*, G) = \frac{w(C)}{|C|}$. 
% The derivative of $g(S, q\cup N_*, G)$ is $g'(S, q\cup N_*, G) = (\frac{3\left|N\right|+\left| N*\right|}{3|N|+\left|N\right|\left|N_*\right|})' = \frac{1}{3|N|+|N|\left|N_*\right|}-\frac{(3|N|+\left|N_*\right|)|N|}{(3|N|+|N|\left|N_*\right|)^2}=\frac{3|N|(1-|N|)}{(3|N|+|N|\left|N_*\right|)^2}$. As $|N| \textgreater 1$, therefore $g'(S, q\cup N_*, G) \textless 0$. Thus $g(S, q\cup N_*, G)$ is monotonically decreasing with regard to $N_*$. Since $|N_*| \geq |N_{opt}|$, the subgraph $q\cup N_{opt}$ contains the query nodes, is connected, and has the highest expected weight gain. Note that $N_{opt}$ is the optimal solution to the set cover problem. Therefore, we can reduce the problem of \searchname~ from the set cover problem.

We set the query nodes as $q = M\cup\{v\} $. To make the return community connected, we need to select some nodes in $\mathcal{N}$. Let $\mathcal{N}_*\subseteq \mathcal{N}$ be a feasible solution for the set cover problem given the element collection $M$ and the set collection $\mathcal{N}$. 
Let $C = q\cup \mathcal{N}_* = M\cup\{v\} \cup N_*$. Then, C is connected, contains the query nodes and has the expected weight gain $g(S, q\cup \mathcal{N}_*, G) = \frac{w(C)}{|V_C|}=\frac{1+\frac{\left|\mathcal{N}_*\right|}{\left|M\right|\left|\mathcal{N}\right|}}{\left|M\right|+1+\left|\mathcal{N}_*\right|}$. 
The derivative of $g(S, q\cup \mathcal{N}_*, G)$ is $g'(S, q\cup \mathcal{N}_*, G) = (\frac{\left|M\right|\left|\mathcal{N}\right|+\left| \mathcal{N}*\right|}{|M|^2|\mathcal{N}|+|M||\mathcal{N}|+|M|\left|\mathcal{N}\right|\left|\mathcal{N}_*\right|})' 
= \frac{1}{|M|^2|\mathcal{N}|+|M||\mathcal{N}|+|M|\left|\mathcal{N}\right|\left|\mathcal{N}_*\right|}-\frac{(|M||\mathcal{N}|+\left|\mathcal{N}_*\right|)|M||\mathcal{N}|}{(|M|^2|\mathcal{N}|+|M||\mathcal{N}|+|M|\left|\mathcal{N}\right|\left|\mathcal{N}_*\right|)^2}$.
Thus, $g'(S, q\cup \mathcal{N}_*, G) =\frac{|M||\mathcal{N}|(|M|+1-|M||\mathcal{N}|)}{(|M|^2|\mathcal{N}|+|M||\mathcal{N}|+|M|\left|\mathcal{N}\right|\left|\mathcal{N}_*\right|)^2}$. As $|M|,|\mathcal{N}| \textgreater 1$, then $|M|+1-|M||\mathcal{N}|\textless 0$ and $g'(S, q\cup \mathcal{N}_*, G) \textless 0$. Thus $g(S, q\cup \mathcal{N}_*, G)$ is monotonically decreasing with regard to $|\mathcal{N}_*|$. Since $|\mathcal{N}_*| \geq |\mathcal{N}_{opt}|$, the subgraph $q\cup \mathcal{N}_{opt}$ contains the query nodes, is connected, and has the highest expected score gain. Note that $\mathcal{N}_{opt}$ is the optimal solution to the set cover problem. Therefore, we can reduce the problem of \searchname~ from the set cover problem.

{\color{black}
Given a community $C$, the expected score gain $g(S, C, G)$ is monotonically decreasing as demonstrated above. With an optimal community $C_{opt}$, its \textit{ESG} is highest. Hence, $|C_{opt}\backslash q|$ is the minimum among feasible solutions where $C_{opt}\backslash q$ are nodes in $C_{opt}$ but not in $q$. According to the graph constructed above, $C_{opt}\backslash q$ is connected to all elements in $M$ and has the minimum size. Thus, the corresponding sets of $C_{opt}\backslash q$ is the optimal solution for the set cover. 

Moreover, by the constructed graph, one node in $G\backslash \{M,v\}$ is mapped to one set in $\mathcal{N}$. Thus, the time complexity for the reduction of the set cover problem and the \textit{IESG} is linear to the node number.
}

% w'(S'\cup q)|S'\cup q|^{-1}-w(S'\cup q)|S'\cup q|^{-2}

{\color{black}\noindent \textbf{Proof of Lemma 2.} For any $\epsilon \textgreater 0$, the set cover cannot be approximated in polynomial time within a ratio of $(1-\epsilon)\ln(|V|)$~\cite{feige1998threshold, DBLP:journals/talg/AlonMS06, DBLP:journals/corr/abs-1305-1979}. We use the reduction from the set cover problem same as the proof of Lemma~1. According to the reduction, community $C$ for the \textit{IESG} problem corresponds to a set collection $\mathcal{N_*}$ for the set cover problem, where each node in $C$ but not $q$ corresponds to one set in $\mathcal{N_*}$. Moreover, the time complexity of reduction is linear to the node number. Thus,
if there is a solution with $(1-\epsilon)\ln(|V|)$-approximation for the \searchname, there will be $(1-\epsilon)\ln(|V|)$-approximation for the optimal solution of set cover, thereby contradicting~\cite{feige1998threshold, DBLP:journals/talg/AlonMS06, DBLP:journals/corr/abs-1305-1979}.}

\noindent \textbf{Proof of Lemma 3.}  $\textit{ESG}(\hat{S}, C_p, G) = \frac{ \mu|S_p|^{\sigma\tau}}{|V_{C_p}|^{\tau}}-\frac{|V_{C_p}|}{|V_{C_p}|^{\tau}}\lambda=\frac{ \mu|V_{C_p}|^{\sigma\tau}}{|V_{C_p}|^{\tau}}-\frac{|V_{C_p}|}{|V_{C_p}|^{\tau}}\lambda$ where $\lambda=\frac{\sum_{u\in V}s_u}{|V|}$. Then the derivative is $\textit{ESG}'(\hat{S}, C_p, G) = \mu(\sigma\tau-\tau)|V_{C_p}|^{\sigma\tau-\tau-1}-(1-\tau)\lambda|V_{C_p}|^{-\tau}$. By setting $\textit{ESG}'(\hat{S}, C_p, G)=0$, we can prove that $\textit{ESG}(\hat{S}, C_p, G)$ increases in $[0, (\frac{(1-\tau)\lambda}{\mu(\sigma\tau-\tau)})^{\frac{1}{\sigma\tau-1}}]$ and decreases in $[(\frac{(1-\tau)\lambda}{\mu(\sigma\tau-\tau)})^{\frac{1}{\sigma\tau-1}}, |V|]$.

\vspace{-2mm}
\subsection{Time Complexity And Extension}
% \vspace{-2mm} 
\noindent \textbf{Time complexity of pre-training.}  The time complexity of the projection of three matrices is $O(3\times(K+1)\times d^2)$. The dot product of query and key takes $O((K+1)^2\times d)$ and the dot product of attention and value also takes $O((K+1)^2\times d)$. Therefore, the self-attention has a time complexity of $O(3\times (K+1)\times d^2 + 2\times(K+1)^2\times d)$. There are $|V|$ nodes in the graph and the number of transformer encoders is $L$, thus the time complexity of \transformername~is $O(L\times|V|\times(3\times (K+1)\times d^2 + 2\times(K+1)^2\times d))$. It is trained for $t$ epochs, therefore, the total time complexity of the offline pre-training is $O(t\times L\times|V|\times(3\times (K+1)\times d^2 + 2\times(K+1)^2\times d))$.

% \vspace{1mm}
% The community score computation computes the pair-wise similarity between the query node representation and graph node representation. 
\noindent \textbf{Time complexity of community score computation.} The time complexity of pair-wise similarity computation is $O(d_m)$ where $d_m$ is the dimension of the latent representation. We need to compute for $|V_q|\times |V|$. Therefore, the overall time complexity of community score computation is $O(|V_q|\times |V|\times d_m)$.

% \vspace{1mm}
% We need to iteratively traverse all the nodes and select the nodes with the highest score. 
\noindent \textbf{Time complexity of \localsearch.} The computation of nodes with the highest score needs $O(|V|log|V|)$ times. We need to compute at most by $|V|$ times. Thus, the overall time complexity of  \localsearch~is  $O(|V|^2log|V|)$.

% \vspace{1mm}
% In \globalsearch, we first sort all the community scores and then select nodes from high community scores to low community scores. 
\noindent \textbf{Time complexity of \globalsearch.} Sorting needs $O(|V|log|V|)$~\cite{cormen2022introduction, ajtai19830}. It iterates at most $log|V|$ iterations, and each iteration needs $O(|V|)$ operations. Therefore, the total time complexity of \globalsearch~is $O(2\times|V|log|V|)$.

\noindent \textbf{Extension and future works.} {\color{black}\framework~is designed for general CS without using labels, and it can be extended to support other related settings of CS. With a specific score computation module to hand additional query input, \framework~can be extended to support attributed community search. With a specific pretraining model, \framework~ can be extended to support other types of graph, e.g., temporal graph. We leave these promising fields in the future works.}

\vspace{-4mm}
\section{Experimental Evaluation}
\label{sec:Experiment}

\begin{table}[t]
\centering %\small %\scriptsize
\caption{Statistics of the datasets}
\vspace{-0.4cm}
\label{tab:datasets}
\begin{tabular} {|p{1.8cm}<{\centering}|p{1.0cm}<{\centering}|p{1.5cm}<{\centering}|p{1.0cm}<{\centering}|p{1.0cm}<{\centering}|}
% {|l|l|l|l|l|}
% {|p{1.8cm}|p{1.0cm}|p{2.0cm}|p{1.0cm}|p{1.0cm}|}
% {|p{3.2cm}|p{4.7cm}|} {|c|c|}
\hline
\cellcolor{lightgray}\textbf{Datasets} & \cellcolor{lightgray}\textbf{$|V|$} & \cellcolor{lightgray}\textbf{$|E|$} & \cellcolor{lightgray}\textbf{$|C|$} & \cellcolor{lightgray}\textbf{$d$}\\ \hline \hline
Texas     & 183     & 325     & 5  & 1,703 \\ \hline
Cornell   & 183     & 298     & 5  & 1,703 \\ \hline
Wisconsin & 251     & 515     & 5  & 1,703 \\ \hline
Cora      & 2,708   & 10,556  & 7  & 1,433 \\ \hline
Citeseer  & 3,327   & 9,104   & 6  & 3,703 \\ \hline
Photo     & 7,650   & 238,162 & 8  & 745   \\ \hline
DBLP      & 17,716  & 105,734 & 4  & 1,639 \\ \hline
CoCS        & 18,333  & 163,788 & 15 & 6,805 \\ \hline
Physics   & 34,493  & 495,924 & 5  & 8,415 \\ \hline
Reddit    & 232,965 & 114,615,892 & 41 & 602  \\ \hline

\end{tabular}
\vspace{-5mm}
\end{table}

\vspace{-1mm}
\subsection{Dataset Description}
% \vspace{-1mm}
% quantities including the total count
We use 10 public datasets from Pytorch Geometric~\cite{Fey/Lenssen/2019} following existing work~\cite{jiang2022query} to comprehensively evaluate the performance. The statistics information of the datasets is summarized in Table~\ref{tab:datasets}. Datasets are characterized by varying numbers of nodes (i.e., $|V|$), numbers of edges (i.e., $|E|$), numbers of communities (i.e., $|C|$) and dimensionalities of features (i.e., $d$).

\vspace{-3mm}
\subsection{Experimental Setup}
% To more comprehensively evaluate \framework,
\noindent\textbf{Baselines:}
We focus on the general community search task. Following~\cite{li2023coclep, jiang2022query}, we compare it with the existing learning-based models including 1) QD-GNN~\cite{jiang2022query}, which is a state-of-the-art supervised method for CS; 2) COCLEP~\cite{li2023coclep}, which is a state-of-the-art semi-supervised method for CS, and the traditional CS methods including 3) CST method~\cite{cui2014local}  that uses \textit{k}-core to model the community; 4) EquiTruss method~\cite{akbas2017truss} that uses \textit{k}-truss to model the community; 5) M$k$ECS method ~\cite{DBLP:conf/cikm/AkibaIY13} that uses \textit{k}-ECC to model the community; and 6) CTC~\cite{huang2015approximate} that aims to find the closest truss community.
\begin{figure}
  \centering
  \includegraphics[width=0.88\linewidth]{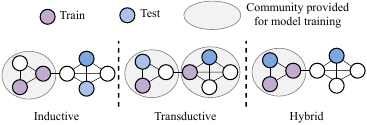}
  \vspace{-4mm}
  \caption{Illustration for the query generation settings}
  \label{fig: query generation}
\vspace{-6mm}
\end{figure}
% [NMI under inductive setting]
% [NMI under transductive setting]
% [NMI under hybrid setting]
% [JAC under inductive setting]
% [JAC under transductive setting]
% [JAC under hybrid setting]
\begin{table*}[t]
\centering %\small %\scriptsize
\caption{F1-score results under different settings}
\vspace{-0.4cm}
\label{tab:exp1_f1score}
\begin{tabular} {p{1.3cm}<{\centering} p{2.2cm}<{\centering} | p{0.8cm}<{\centering} p{0.8cm}<{\centering} p{1.0cm}<{\centering} p{0.8cm}<{\centering} p{0.8cm}<{\centering} p{0.8cm}<{\centering} p{0.8cm}<{\centering} p{0.8cm}<{\centering} p{0.8cm}<{\centering} p{0.8cm}<{\centering} | p{1.5cm}<{\centering}}
% {|l|l|l|l|l|}
% {|p{1.8cm}|p{1.0cm}|p{2.0cm}|p{1.0cm}|p{1.0cm}|}
% {|p{3.2cm}|p{4.7cm}|} {|c|c|}
\toprule
\textbf{Settings} & \textbf{Models} & Texas & Cornell & Wisconsin & Cora& Citeseer& Photo& DBLP& CoCS & Physics & Reddit & Average +/- \\ \hline \hline
\multirow{8}{*}{\textbf{Inductive}} 
& CST     & 0.1986 & 0.1975 & 0.2251 & 0.2111 & 0.1423 & 0.2019 & 0.2854 & 0.1252 & 0.2276 & 0.1463  & -27.12\%  \\ 
& EquiTruss   & 0.3120 & 0.3168 & 0.3079 & 0.2384 & 0.2240  & 0.2166 & 0.3252 & 0.1225 & 0.2471 & 0.2163 & -21.46\% \\ 
& M$k$ECS & 0.3581 & 0.3177 & \underline{0.3404} & 0.2364 & 0.2015 & 0.1975 & 0.2768 & 0.1152 & 0.2193 & 0.2068 & -22.03\%  \\ 
& CTC    & 0.3211  & \underline{0.3482} & 0.3327 & 0.2558 & 0.2418 & 0.2626 & 0.3417 & 0.1059 & 0.2511 & 0.2431 & -19.69\%  \\ 
& QD-GNN     & 0.0821 & 0.0669 & 0.0683 & 0.0322 & 0.0536 & 0.0018 & 0.0372 & 0.0145 & OOM    & OOM  & -41.50\%  \\ 
& COCLEP  & \underline{0.4044} & 0.2960 & 0.1804 & 0.3094 & 0.3058 & 0.4413 & 0.3066 & 0.4253 & 0.3389 & 0.2696 & -13.95\%\\ 
& \framework-LS     & 0.1801 & 0.1583 & 0.2074 & \underline{0.5467} & \underline{0.3906} & \underline{0.5725} & \textbf{0.4407} & \underline{0.4292} & \underline{0.5075} & \textbf{0.4879} & -7.52\%\\ 
& \framework-GS     & \textbf{0.4283} & \textbf{0.3716} & \textbf{0.3755} & \textbf{0.5764} & \textbf{0.4535} & \textbf{0.6018} & \underline{0.4326} & \textbf{0.4374} & \textbf{0.5113} & \underline{0.4848} & - \\ \midrule

\multirow{2}{*}{\textbf{Transductive}} 
& QD-GNN    & 0.6703 & 0.8408 & 0.6247 & 0.5062 & 0.4726 & 0.2205 & 0.4918 & 0.6356 & OOM    & OOM & +9.81\%   \\ & COCLEP    & 0.4020 & 0.3167 & 0.3206 & 0.3685 & 0.3331 & 0.5060 & 0.3763 & 0.3549 & 0.4388 & 0.3270 & -9.29\% \\ 
\midrule

\multirow{2}{*}{\textbf{Hybrid}} 
& QD-GNN    & 0.3852 & 0.3644 & 0.5956 & 0.4789 & 0.4097 & 0.0833 & 0.3902 & 0.4969 & OOM    & OOM  & -5.91\%  \\ 
& COCLEP   & 0.3883 & 0.3313 & 0.2938 & 0.3615 & 0.3067 & 0.4388 & 0.3733 & 0.4027 & 0.4693 & 0.3071 & -10.01\%  \\ 

\bottomrule

\end{tabular}
    % \vspace{-1mm}
    \begin{flushleft}
        \footnotesize $*$ CST, EquiTruss, M$k$ECS, CTC and \framework~have consistent results under three settings as they are label-free.   \framework~with \localsearch~is denoted as  \framework-LS, and \framework~with \globalsearch~is denoted as  \framework-GS. OOM indicates out-of-memory. The last column presents the average margin compared to \framework-GS. 
    \end{flushleft} 
\vspace{-5mm}
\end{table*}

\begin{figure*}
\subfigbottomskip=-2pt %设置第二行子图与第一行子图的距离，即下面的头与上面的脚的距离
\subfigcapskip=-6pt %设置子图与子标题之间的距离
    
    \subfigure{ 
        
        \includegraphics[width=0.32\textwidth]{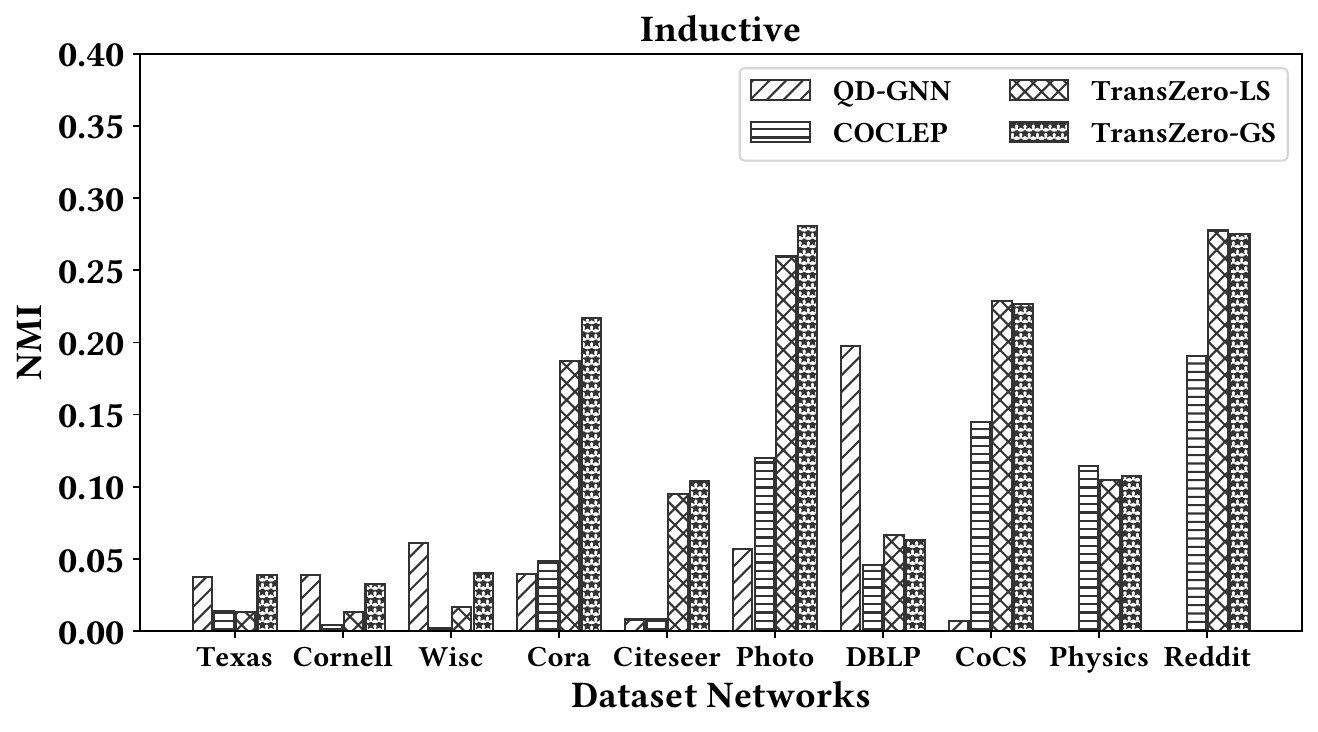}
    \vspace{-10mm}
    }
    \vspace{-3mm}
    \subfigure{ 
        
        \includegraphics[width=0.32\textwidth]{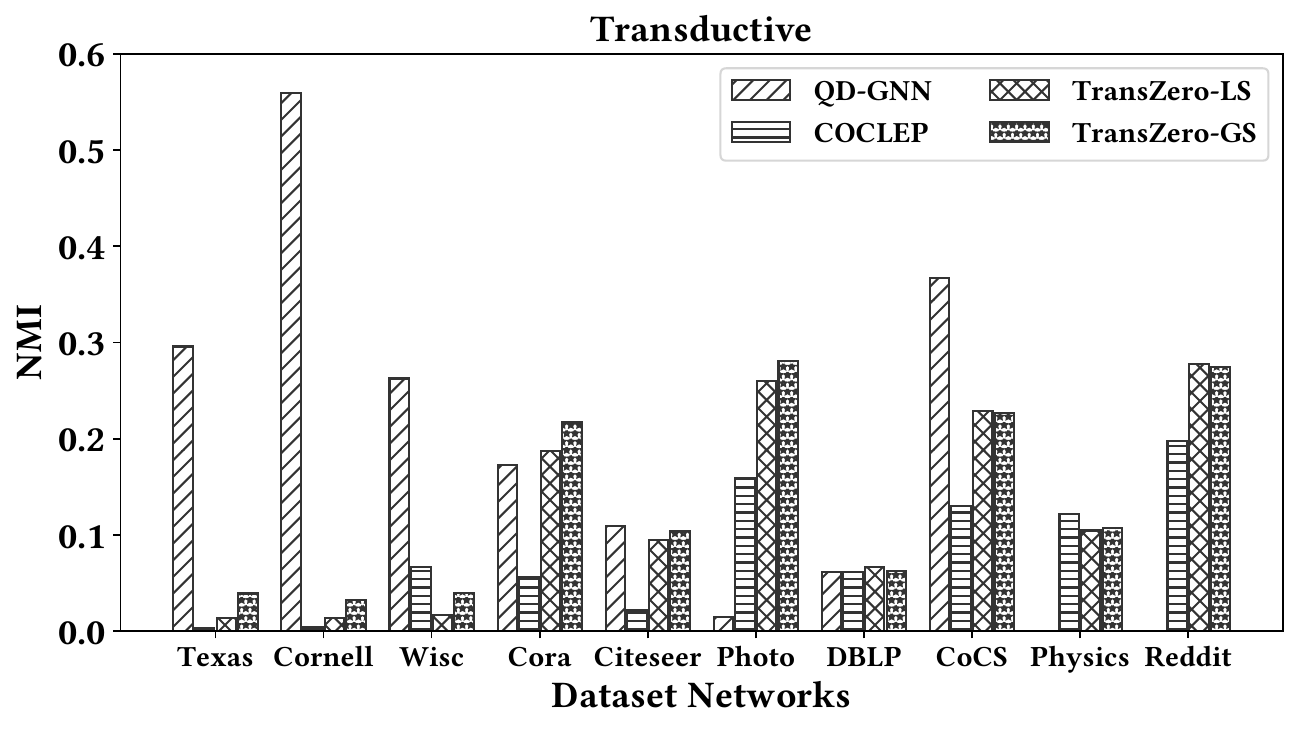}
    \vspace{-10mm}
    }
    \subfigure{
        % \centering
        \includegraphics[width=0.32\textwidth]{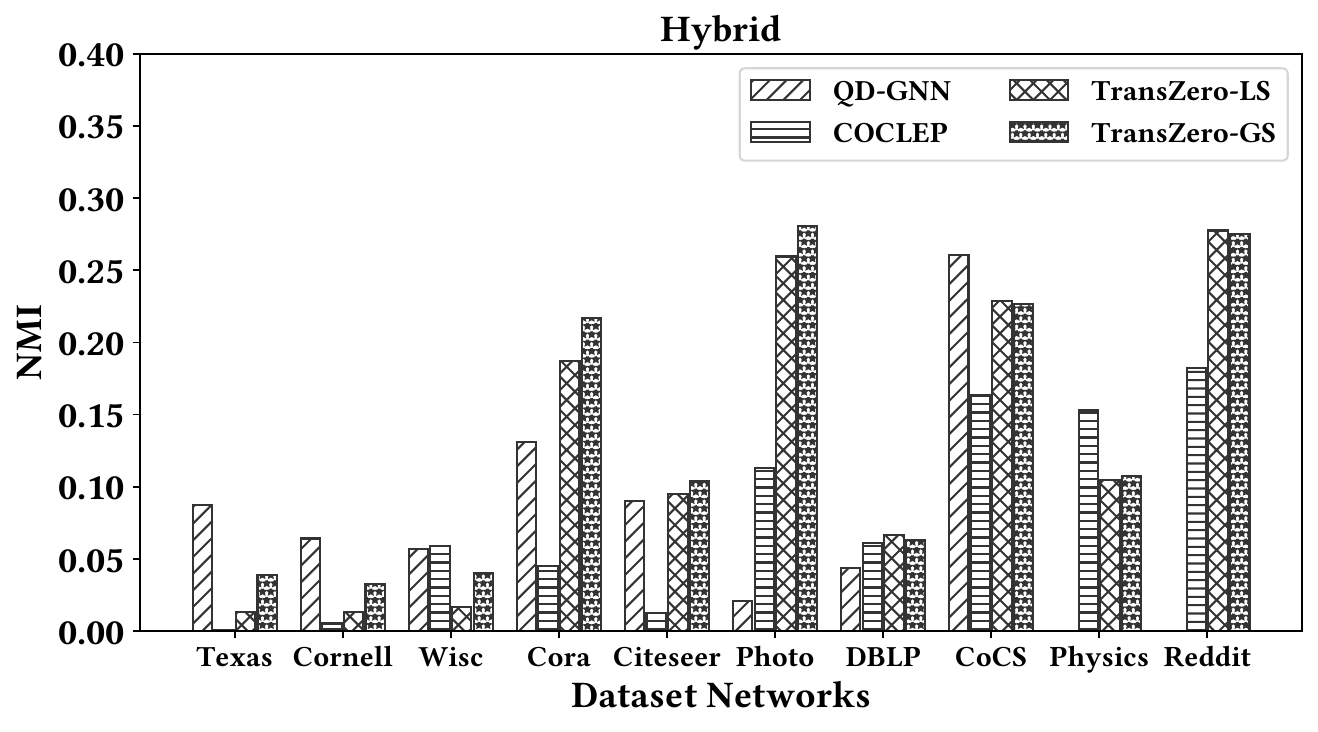} 
    \vspace{-10mm}
    }
    \subfigure{
        \vspace{-5mm}
        \includegraphics[width=0.32\textwidth]{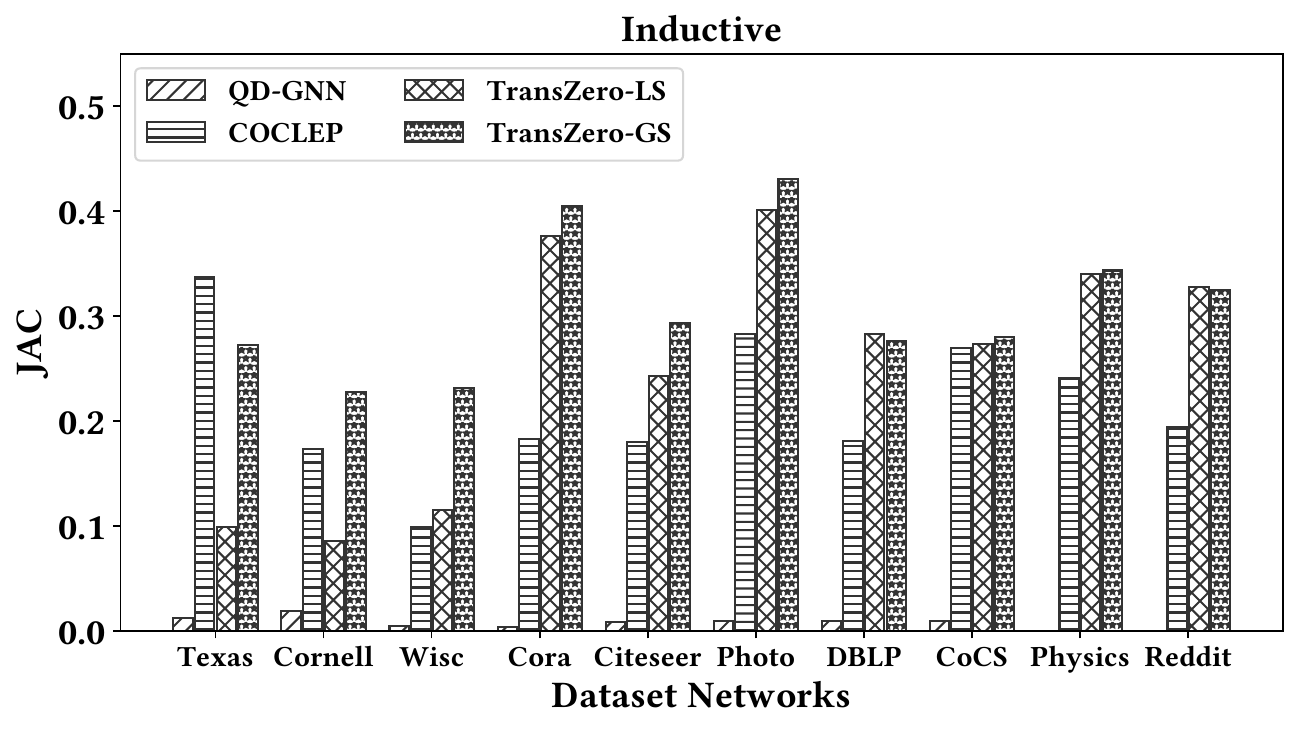} 
    \vspace{-3mm}
    }
    \subfigure{
        \vspace{-5mm}
        \includegraphics[width=0.32\textwidth]{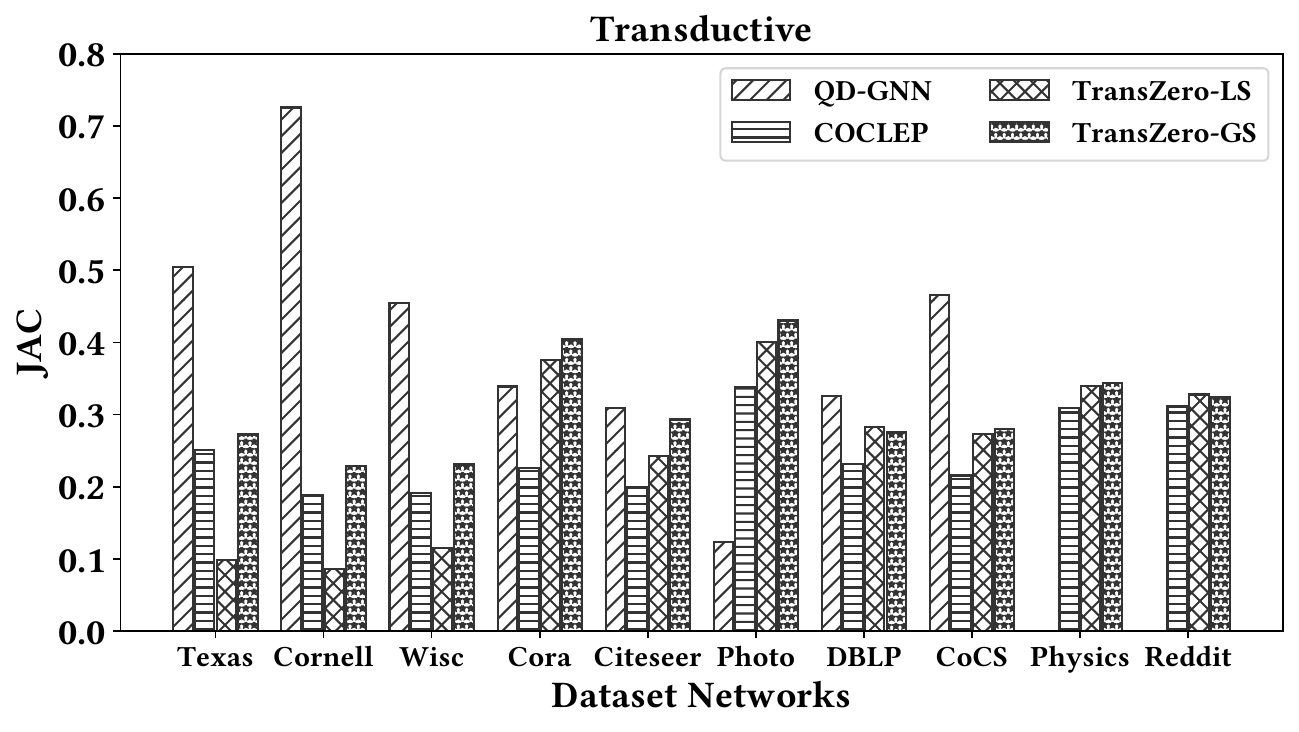} 
    \vspace{-3mm}
    }
    % \vspace{-3mm}
    \subfigure{ 
        \vspace{-5mm}
        \includegraphics[width=0.32\textwidth]{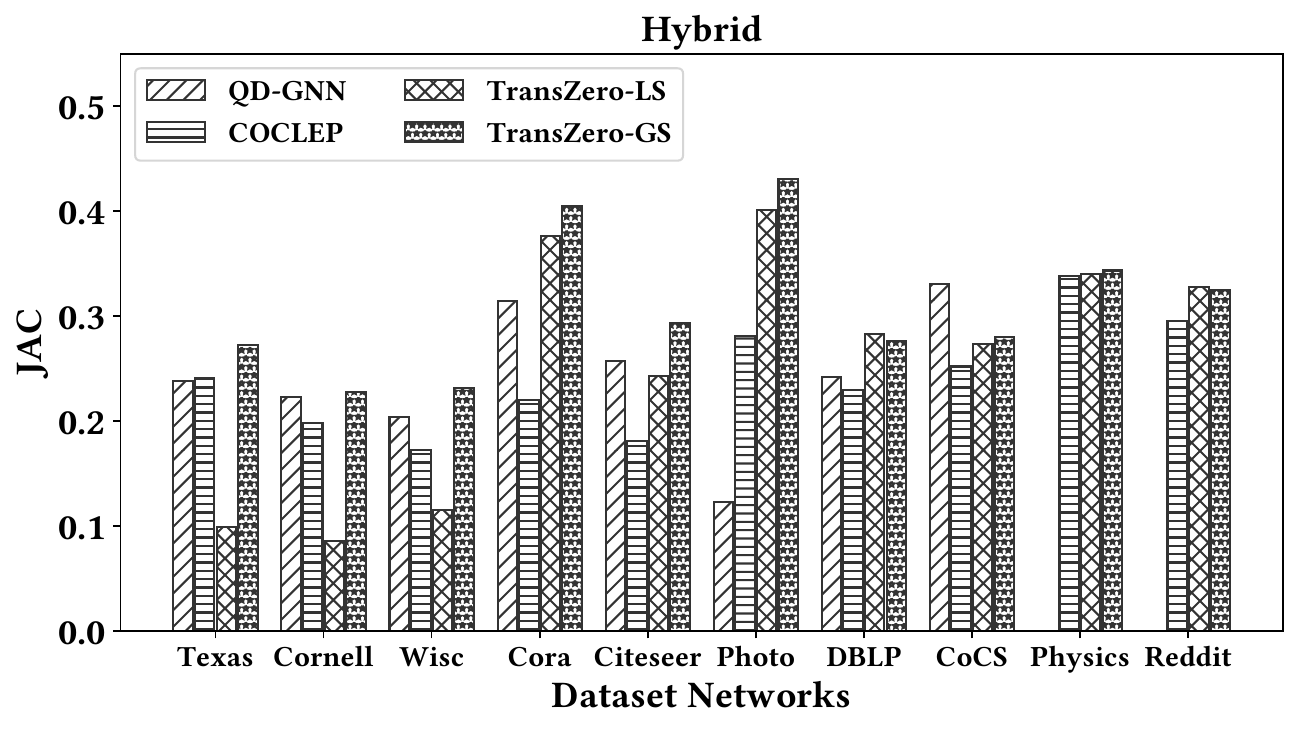}
    \vspace{-3mm}
    }
\vspace{-5mm}
    \caption{NMI and JAC results under different settings}
    \label{fig:nmi_jac_results}
\vspace{-6mm}
\end{figure*}

\vspace{1mm}
\noindent\textbf{Query generation:}
We use the following three generation mechanisms to generate queries for training, validation and testing.

\begin{itemize}[leftmargin=10pt, topsep=1pt]
\item{} \textbf{Inductive Setting.} We randomly partition all ground-truth communities into two groups including training communities and testing communities with a ratio of about 1: 1. And we generate training and validation queries from the training communities and generate test queries from the testing communities. This setting aims to test the ability to predict unseen communities. 
\item{} \textbf{Transductive Setting.} We generate all the queries randomly from all the ground-truth communities.
\item{} \textbf{Hybrid Setting.} We randomly divide ground-truth communities into training and testing groups ($\sim$1:1 ratio). Training and validation queries are generated from the training communities, while test queries are generated from all ground-truth communities. This setting closely simulates real-world scenarios by training on a subset of known ground-truth communities and evaluating across all the ground-truth communities.
\end{itemize}
An illustration of the above three settings is shown in Figure~\ref{fig: query generation}. Note that QD-GNN is evaluated in a transductive manner in the original paper, and COCLEP is evaluated in a transductive manner when the number of ground-truth communities is small and in an inductive manner when the number of ground-truth communities is large as in the original paper. It is worth noting that methods that do not need ground-truth communities (i.e.,\framework~and the traditional CS methods) exhibit consistent performance across different generation settings. 
Consistent with ~\cite{jiang2022query}, the number of training queries, validation queries and testing queries are 150, 100 and 100, respectively.
Following~\cite{jiang2022query}, we randomly select 1 to 3 nodes from the ground-truth community as the query nodes. As in the original paper of COCLEP~\cite{li2023coclep}, we generate 3 positive samples beside the query node for COCLEP.

\noindent\textbf{Metrics:} 
In this paper, we mainly focus on F1-score~\cite{sasaki2007truth} that is commonly used by existing works~\cite{li2023coclep, jiang2022query} to evaluate the quality of the found community. Besides the F1-score, we also utilize Normalized Mutual Information (NMI) ~\cite{danon2005comparing} and Jaccard similarity (JAC)~\cite{zhang2020seal} aligned with COCLEP~\cite{li2023coclep} for evaluation. We follow the calculation of F1-score used in ~\cite{jiang2022query}. For all the F1-score, NMI and Jaccard, a higher value indicates a better found community. 

\noindent\textbf{Implementation Details:} We run \framework~for 100 epoches with early stopping. The maximum number of hops used in the augmented subgraph sampler is 5. The value of $\tau$ is set as $0.5$ for all datasets. The value of $\alpha$ is set as 0.1 for all datasets. The dropout rate is set as $0.1$. The batch size is set as the number of nodes in the graph or 4000 if it runs out of memory. The number of attention heads is 8. We limit our search to a maximum of 50\% of the total nodes and do not exceed 10,000 nodes. The number of transformer layer is the same with ~\cite{chen2022nagphormer}.  \globalsearch~is used as the default online search method for \framework. The hyper-parameters of QD-GNN and COCLEP are the same as in their original paper. Experiments are conducted on a server with Intel(R) Xeon(R) Gold 6342 CPU, 503GB memory and Nvidia RTX 4090 (GPU).

\begin{figure*}
\subfigcapskip=-7pt %设置子图与子标题之间的距离
    
    \subfigure[Efficiency results of the training phase]{ 
        
        \includegraphics[width=0.425\textwidth]{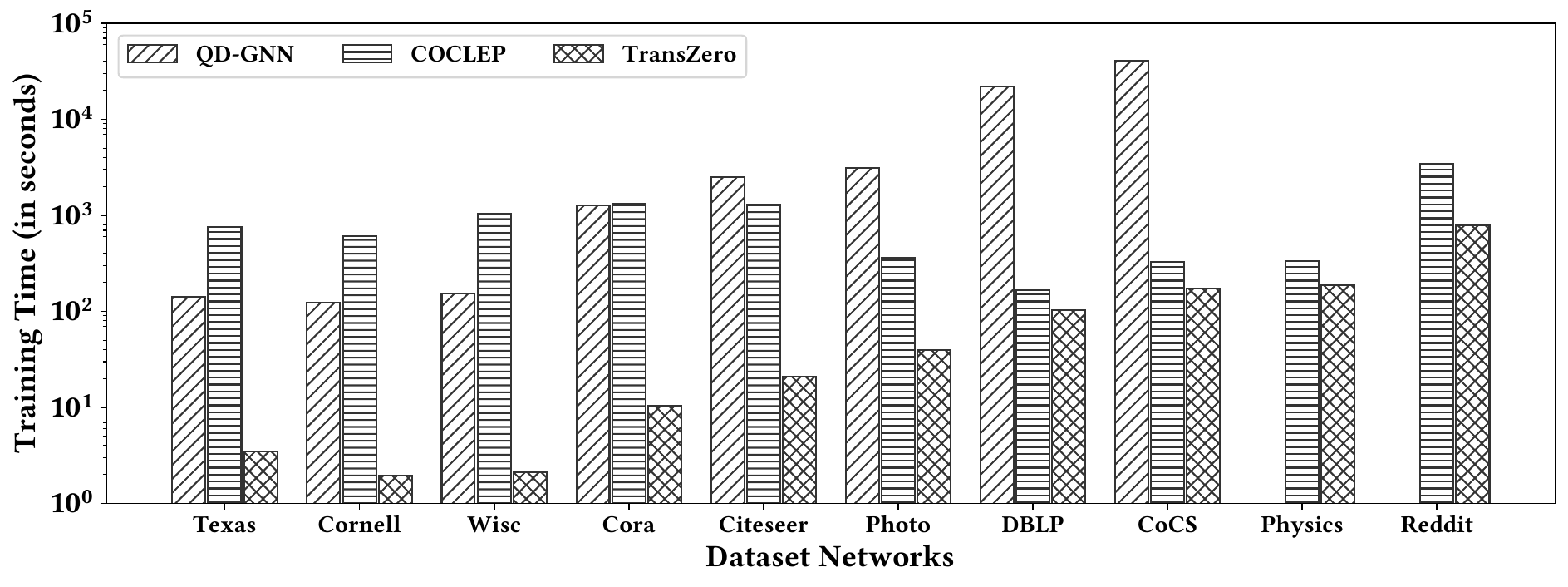}
    % \vspace{-3mm}
    }
    \subfigure[Efficiency results of the search phase]{ 
        \includegraphics[width=0.425\textwidth]{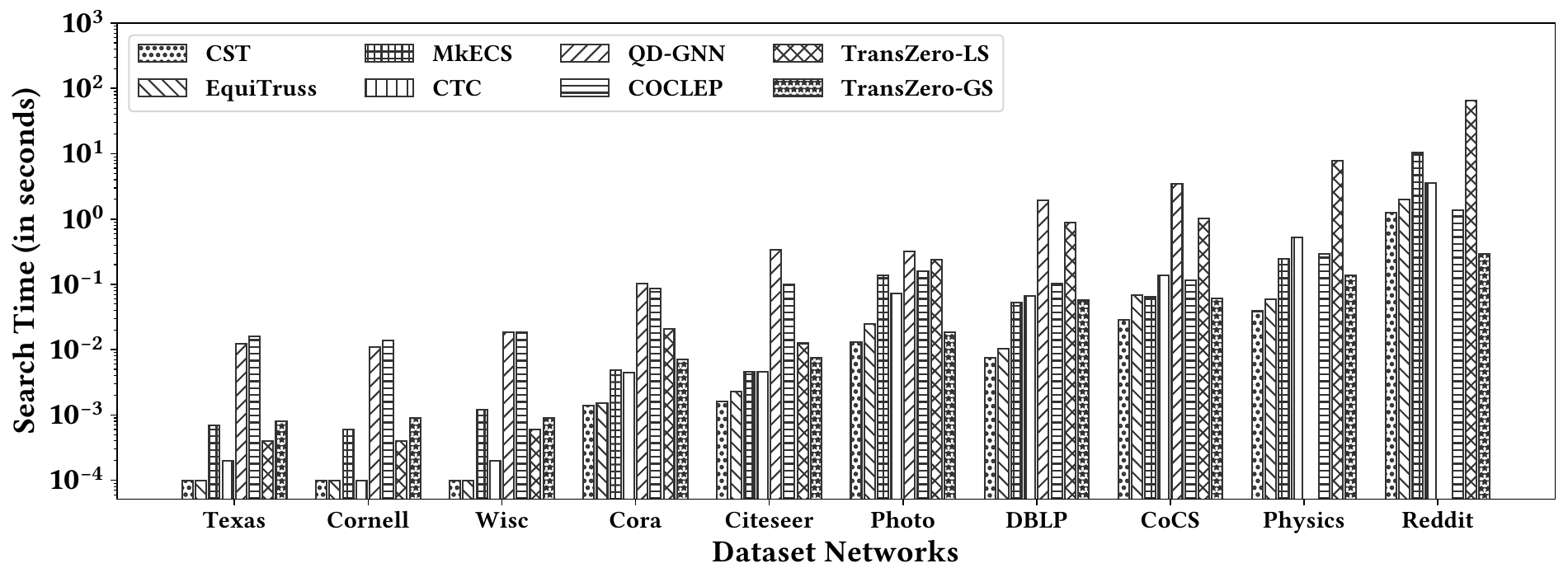}
    % \vspace{-3mm}
    }
\vspace{-6mm}
    \caption{{\color{black}Efficiency results}}
    \label{fig:efficiency_evaluation}
\vspace{-6mm}
\end{figure*}

\begin{figure*}
\subfigbottomskip=-3pt %设置第二行子图与第一行子图的距离，即下面的头与上面的脚的距离
\subfigcapskip=-7pt %设置子图与子标题之间的距离
    
    \subfigure[F1-score with varying $\alpha$]{ 
        
        \includegraphics[width=0.32\textwidth]{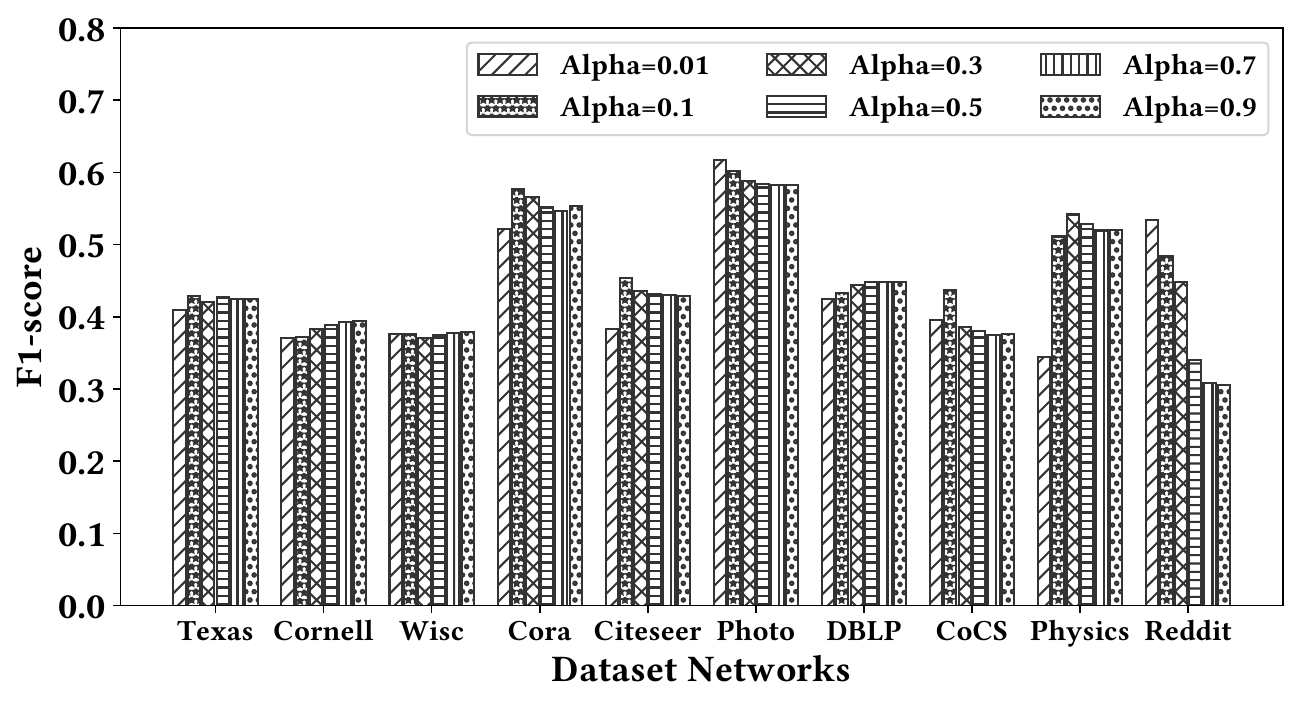}
    \vspace{-3mm}
    }
    \vspace{-1mm}
    \subfigure[F1-score with varying $\tau$]{ 
        
        \includegraphics[width=0.32\textwidth]{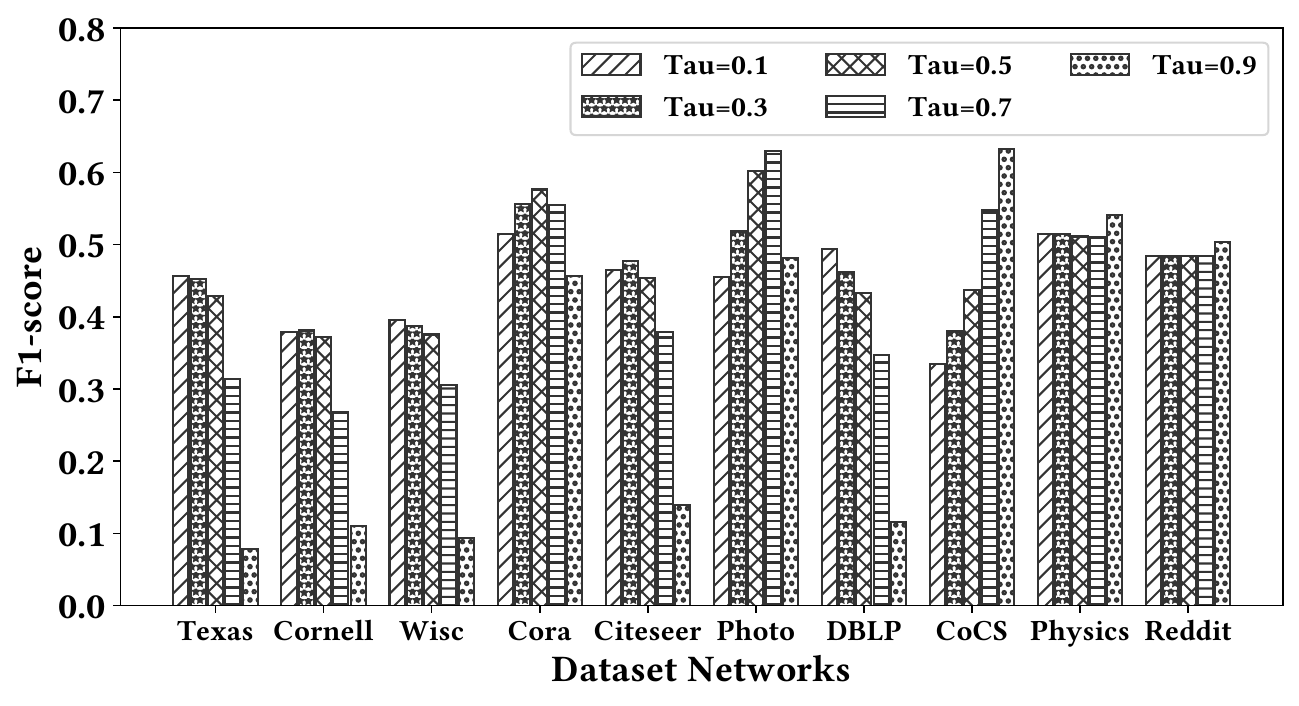}
    \vspace{-3mm}
    }
    \subfigure[F1-score with varying similarity definitions]{
        % \centering
        \includegraphics[width=0.32\textwidth]{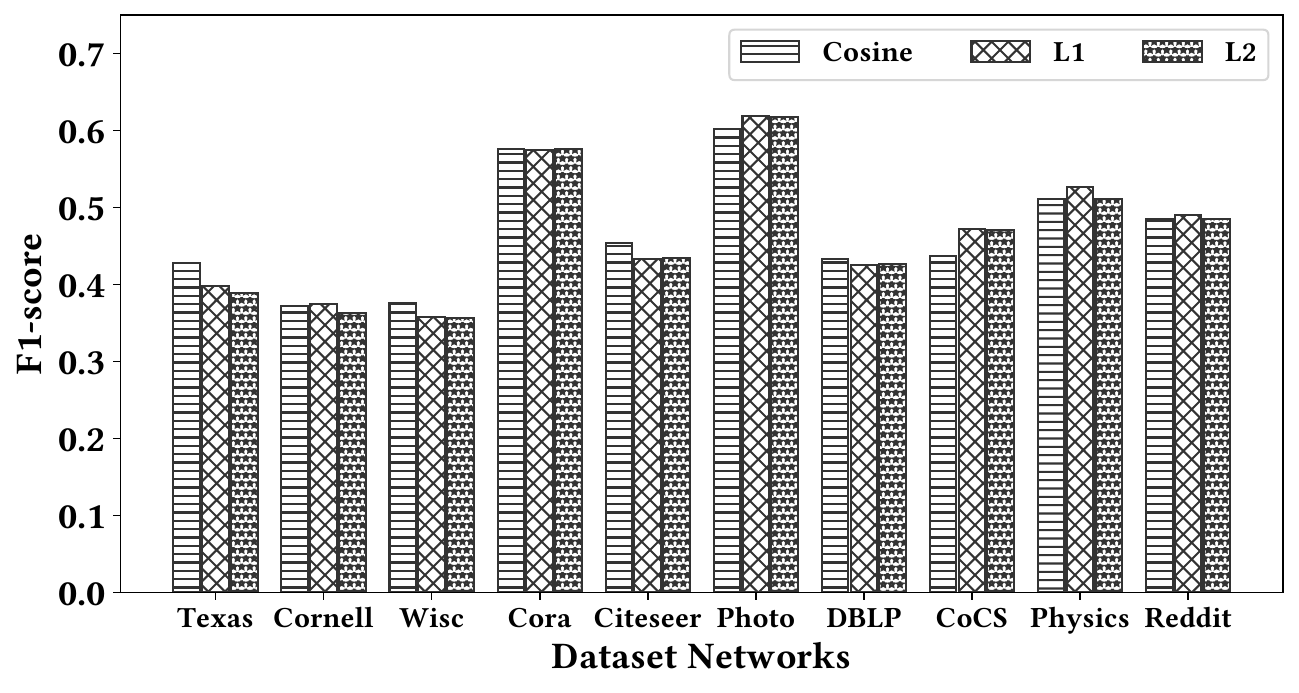} 
    \vspace{-3mm}
    }
    \subfigure[F1-score with varying hop numbers]{
        \vspace{-5mm}
        \includegraphics[width=0.32\textwidth]{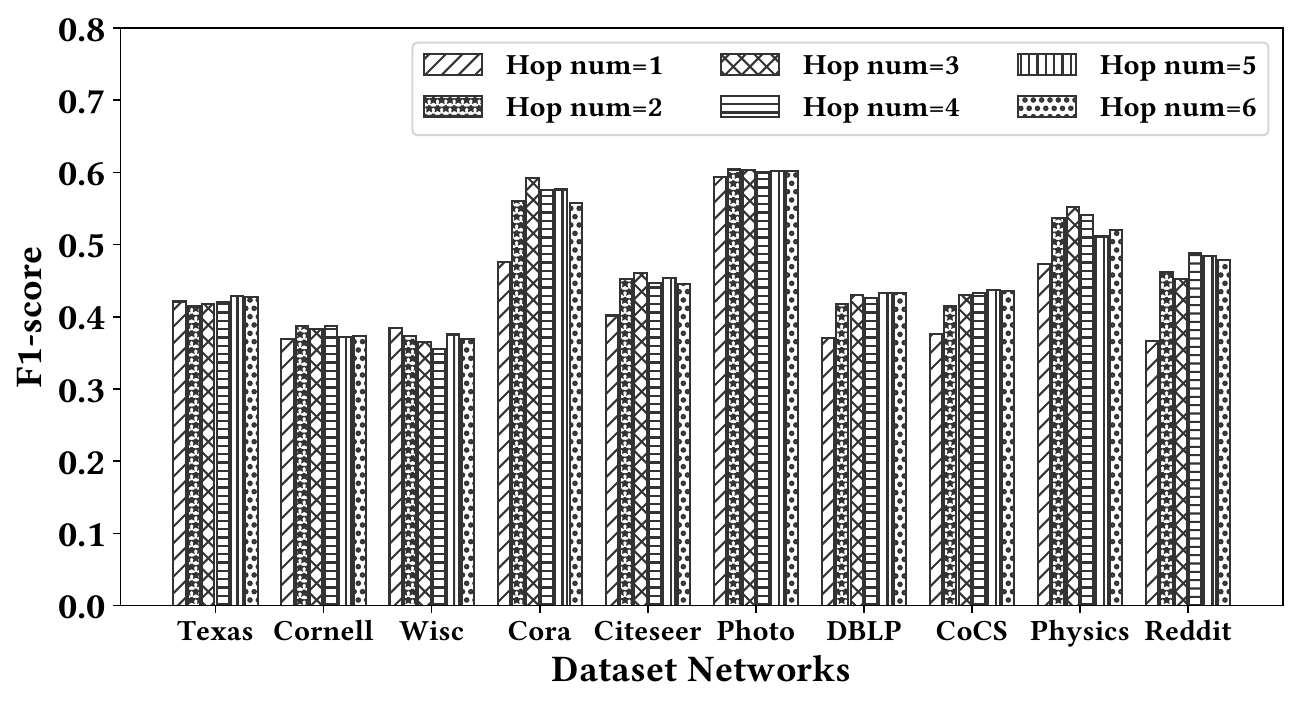} 
    \vspace{-3mm}
    }
    \subfigure[F1-score with varying epoch numbers]{
        \vspace{-5mm}
        \includegraphics[width=0.32\textwidth]{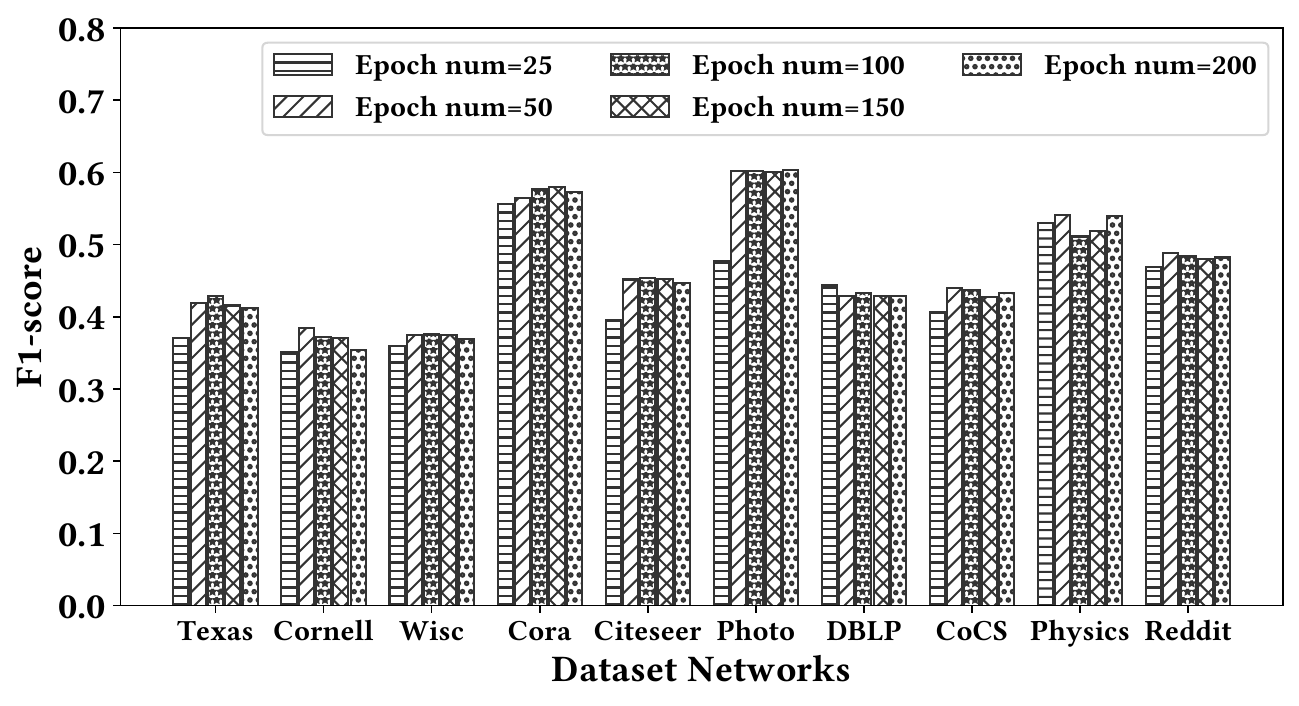} 
    \vspace{-3mm}
    }
    \subfigure[F1-score with varying identification strategies]{ 
        \vspace{-5mm}
        \includegraphics[width=0.32\textwidth]{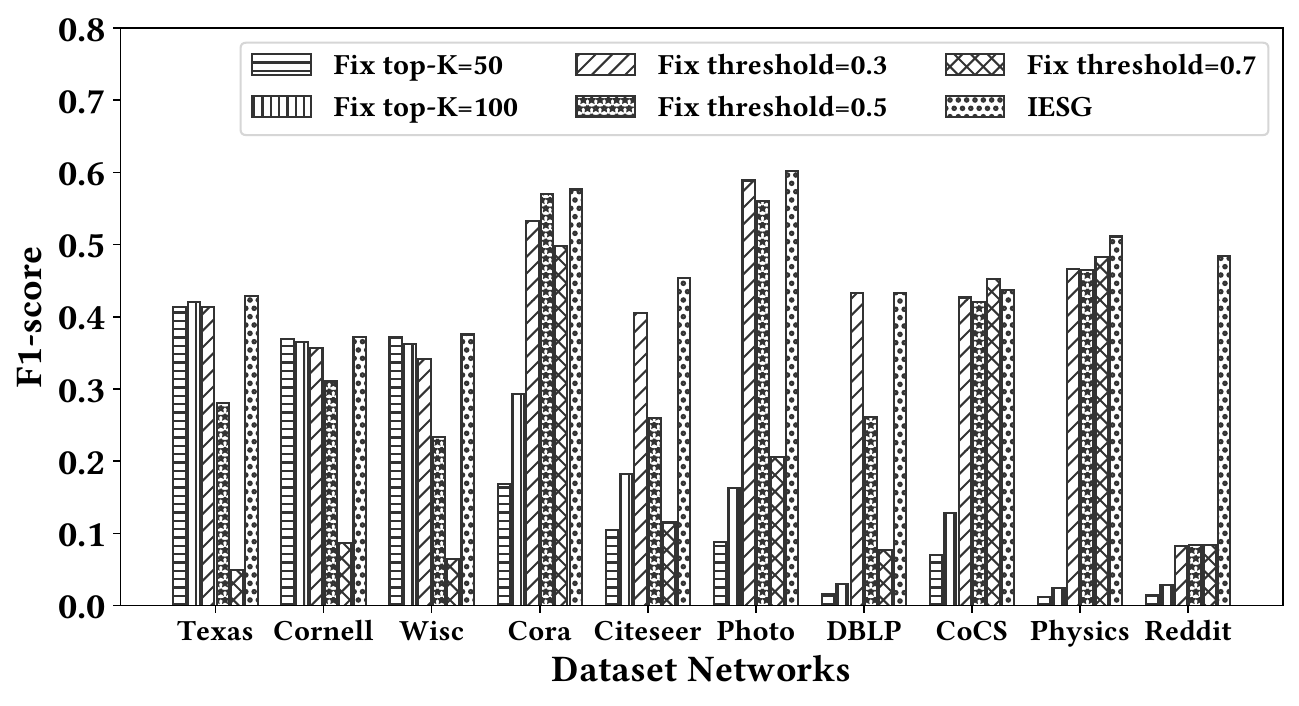}
    \vspace{-3mm}
    }
\vspace{-4mm}
    \caption{Hyper-parameter analysis results}
    \label{fig:hyperparameter_analysis}
\vspace{-6mm}
\end{figure*}

\vspace{-5mm}
\subsection{Effectiveness Evaluation}
\vspace{-1mm}
\noindent\textbf{Exp-1: F1-score results.} We first present the F1-score results across three settings in Table~\ref{tab:exp1_f1score}. Note that traditional CS methods and \framework~have consistent results across three settings as they are label-free, and we solely present their performance in the inductive setting to avoid redundancy. \framework~with \localsearch~(\textit{resp.} \globalsearch) is denoted as  \framework-LS~(\textit{resp.} \framework-GS). OOM indicates out-of-memory. 
Among the traditional methods, both EquiTruss and CTC demonstrate competitive performances, and \framework-GS~outperforms EquiTruss by 21.46\% and surpasses CTC by 19.69\%.
Among the learning-based models, the performance of QD-GNN varies significantly across settings. QD-GNN needs all nodes in the ground-truth communities for training and thus good at the transductive setting as it can memorize all the communities. However, it is hard to generalize its performance to settings with unseen communities (i.e., inductive setting and hybrid setting).
In the inductive setting, \framework-GS significantly outperforms QD-GNN and COCLEP with an average F1-score enhancement of 41.50\% and 13.95\%, respectively.
In the hybrid setting, \framework-GS exhibits outstanding performance, outperforming QD-GNN and COCLEP by an average F1-score of 5.91\% and 10.01\%, respectively. These results highlight the effectiveness of \framework~measured by F1-score.

\vspace{1mm}
\noindent\textbf{Exp-2: NMI and JAC results under different settings.} In this part, we use the NMI and JAC to measure the learning-based methods as their high performance demonstrated in Exp-1. The results are presented in Figure~\ref{fig:nmi_jac_results}. The figure shows that QD-GNN still has the best performance under the transductive setting, and the performance is hard to generalize to the inductive setting and the hybrid setting. In terms of NMI, \framework-GS outperforms QD-GNN with an average enhancement of 9.75\% and surpasses COCLEP by 5.86\% under the hybrid setting. In terms of JAC, \framework-GS outperforms QD-GNN by an average of 3.27\% and surpasses COCLEP by 6.75\% under the hybrid setting. These results demonstrate the high performance of \framework~regarding both NMI and JAC.

\vspace{-4mm}
\subsection{Efficiency Evaluation}
\vspace{-1mm}

\noindent\textbf{Exp-3: Efficiency evaluation.}  In Figure~\ref{fig:efficiency_evaluation}, we report the efficiency results, including the efficiency of both the training phase and the search phase. 
{\color{black}Note that there is no training phase for the traditional CS methods, and thus we only report their efficiency results of the search phase}.
In terms of the training phase, \framework~significantly outperforms the existing learning-based methods. It achieves an average speedup of 118.22$\times$ and up to 235.83$\times$ in dataset CoCS compared to QD-GNN. When compared to COCLEP, \framework~achieves an average speedup of 122.39$\times$ and up to 486.07$\times$ in Wisconsin. 
Regarding the search phase,
\framework-GS achieves an average speedup of 26.77$\times$ and reaches up to 56.48$\times$ in the CoCS dataset when compared to QD-GNN. When compared to COCLEP, \framework~achieves an average speedup of 10.02$\times$ and up to 20.41$\times$ in Wisconsin.  Moreover, \framework-LS has a better performance on small datasets. 
{\color{black}Compared to the traditional CS methods, \framework-GS also shows a competitive performance, particularly on large datasets}.
The results validate the superior efficiency of \framework.

\vspace{-4mm}
\subsection{Hyper-parameter Analysis}
\vspace{-1mm}
\noindent\textbf{Exp-4: Varying $\alpha$.}
In Figure~\ref{fig:hyperparameter_analysis}(a), we evaluate \framework~with varying values of $\alpha$. Note that $\alpha$ is utilized to balance the personalization loss and the link loss in Equation~\ref{equ:loss_func}. We set the value equal to 0.01, 0.1, 0.3, 0.5, 0.7 and 0.9 respectively. The figure shows that the value of $\alpha$ has a different impact on different datasets. In Cornell and DBLP, the performance of \framework~improves with the increase of $\alpha$. On the datasets like Reddit and Photo, the performance decreases with the increase of  $\alpha$. In general, $\alpha$ with a value of 0.1 could effectively balance the two losses and achieve a good performance.

\begin{table*}[t]
\centering %\small %\scriptsize
\caption{Ablation study}
\vspace{-0.4cm}
\label{tab:exp10_ablation}
\resizebox{\linewidth}{!}{
\begin{tabular} {p{3.2cm}<{\centering} |p{0.9cm}<{\centering} p{0.9cm}<{\centering} p{1.0cm}<{\centering} p{0.9cm}<{\centering} p{0.9cm}<{\centering} p{0.9cm}<{\centering} p{0.9cm}<{\centering} p{0.9cm}<{\centering} p{0.9cm}<{\centering} p{0.9cm}<{\centering}|  p{1.8cm}<{\centering}}
% {|l|l|l|l|l|}
% {|p{1.8cm}|p{1.0cm}|p{2.0cm}|p{1.0cm}|p{1.0cm}|}
% {|p{3.2cm}|p{4.7cm}|} {|c|c|}
\toprule
\textbf{Models} & Texas & Cornell & Wisconsin & Cora& Citeseer& Photo& DBLP& CoCS & Physics & Reddit & Average $+/-$ \\ \hline \hline

Full model  & 0.4283 & 0.3716 & 0.3755 & 0.5764 & 0.4535 & 0.6018 & 0.4326 & 0.4374 & 0.5113 & 0.4848 & - \\
w/o $\mathcal{L}_p$     & 0.4215 & 0.3749 & 0.3773 & 0.5462 & 0.4259 & 0.5716 & 0.4501 & 0.3502 & 0.5183 & 0.2981 & -3.19\% \\
w/o $\mathcal{L}_k$ & 0.3894 & 0.3576 & 0.3579 & 0.4203 & 0.3044 & 0.6116 & 0.4087 & 0.4532 & 0.3506 & 0.5076 & -5.12\%\\
\text{\color{black}w/o Conductance Aug }& \text{\color{black} 0.4212 } & \text{\color{black} 0.3692 } &	\text{\color{black} 0.3848 }  &	\text{\color{black} 0.4755 }  & \text{\color{black} 	0.4019 } &\text{\color{black} 0.5935 }	 &\text{\color{black} 0.3708 }	 &\text{\color{black} 0.3766 }	&\text{\color{black} 0.4738 }	 &\text{\color{black} 0.4167 }	 & \text{\color{black} -3.89\% }\\
w/o \transformername & 0.3317 & 0.2421 & 0.2169 & 0.4048 & 0.2780  & 0.4473 & 0.2708 & 0.3074 & 0.3435 & 0.3649 & -14.65\%\\ 

\bottomrule
\end{tabular}
}
\vspace{-4mm}
\end{table*}

%  granularity of the subgraph, and a higher $\tau$ value leads to a more fine-grained 
% \vspace{1mm}
\noindent\textbf{Exp-5: Varying $\tau$.} In Figure~\ref{fig:hyperparameter_analysis}(b), we evaluate \framework~across various values of $\tau$. $\tau$ is utilized to regulate the granularity of the subgraph, as defined in Definition\ref{definition:esg}, with higher $\tau$ resulting in a more fine-grained subgraph. We consider $\tau$ values of 0.1, 0.3, 0.5, 0.7, and 0.9.
The results show that the performance of \framework~experiences a decrease with the increase of $\tau$ for smaller datasets such as Texas. Conversely, for datasets like CoCS and Reddit, the performance of \framework~improves with higher $\tau$ values. In general, $\tau$ with a value of 0.5 consistently demonstrates excellent performance.

% he L1-similarity is defined as $1-\sum_{i=0}^{d_M^{(L)}}|f^{\theta}_i(v)-f^{\theta}_i(u)|$, and the L2-similarity is defined as $1-\sqrt{\sum_{i=0}^{d_M^{(L)}}(f^{\theta}_i(v)-f^{\theta}_i(u))^2}$. 
% \vspace{1mm}
\noindent\textbf{Exp-6: Varying similarity definitions.} In this part, we evaluate \framework~using different similarity definitions. We replace line 4 in Algorithm~\ref{algo:score_computation} with L1-similarity and L2-similarity. The results are in Figure~\ref{fig:hyperparameter_analysis}(c). The results show that the performance using different similarities achieves a close performance. The results validate the robustness of \framework~to different similarity definitions.

% \vspace{1mm}
\noindent\textbf{Exp-7: Varying hop numbers.} In Figure~\ref{fig:hyperparameter_analysis}(d), we evaluate \framework~across varying numbers of hops, which are employed in the augmented subgraph sampler. We consider the number of hops as 1, 2, 3, 4, 5, and 6, respectively. 
The results show that 1 hop information proves sufficient for small graphs such as Wisconsin. Conversely, for medium and large graphs, a larger number of hops is necessary. For instance, in the case of Reddit, \framework~achieves optimal performance with 4 hops and 5 hops. In summary, 5 hops are generally sufficient for \framework~to achieve optimal performance.

% \vspace{1mm}
\noindent\textbf{Exp-8: Varying epoch numbers.} In Figure~\ref{fig:hyperparameter_analysis}(e), we present the performance of \framework~across varying numbers of training epochs during the pre-training phase. We consider the number of epochs as 25, 50, 100, 150, and 200, respectively.
As depicted in the figure, the performance exhibits similarity after 50 epochs, suggesting that the model has reached convergence within the first 50 epochs, and 100 epochs prove to be sufficient for effective model training.

% \vspace{1mm}
% \noindent\textbf{Exp-9: Coverage analysis.} In this part, we provide an analysis of loss trends and coverage patterns. We illustrate the evolution of loss across different epochs for all datasets in Figure \ref{fig:hyperparameter_analysis}(f). Our findings reveal that certain datasets, such as Reddit and Physics, achieve early coverage, while others, like Photo, require more epochs to achieve satisfactory coverage. For instance, datasets like Reddit exhibit convergence at approximately 30 epochs and early stops at about 50 epochs. Generally, most datasets experience a significant reduction in loss within the initial 40 epochs, and they tend to achieve coverage at around 40 epochs.

% the strategy employing a fixed number for community identification performs well on small datasets but poorly on large datasets. Conversely, the strategy using a fixed threshold for community identification performs well on small datasets but poorly on large datasets.

% \vspace{1mm}
\noindent\textbf{Exp-9: Varying identification strategies.} 
In Figure \ref{fig:hyperparameter_analysis}(f), we present the results of various identification strategies discussed in Section~\ref{sec:Introduction}. We compare the fixed number strategy and the fixed threshold strategy. The fixed number is set as 50 and 100, and the fixed threshold is set as 0.3, 0.5, and 0.7, respectively.
The figure indicates that neither the fixed-number-based strategy nor the fixed-threshold-based strategy generalizes well across all evaluated datasets. In contrast, our proposed \searchname~consistently demonstrates strong performance. These findings highlight the effectiveness of our proposed \searchname.

 % which is designed specifically for the task of CS
% consistently contributes to the performance enhancement of the \framework
\vspace{-6mm}
\subsection{Ablation Study and Case Study}
\vspace{-1mm}
\noindent\textbf{Exp-10: Ablation study.}
In this section, we investigate the effectiveness of components employed in \framework, including the personalized loss $\mathcal{L}_p$, the link loss $\mathcal{L}_k$, the conductance-based subgraph sampler and the \transformername.  The results are presented in Table~\ref{tab:exp10_ablation}.
Regarding the personalization loss, its impact becomes apparent in scenarios involving medium and large graphs. Especially, when applied to the Reddit dataset, $\mathcal{L}_p$ exhibits a remarkable enhancement in the F1-score, enabling an F1-score increase of 18.67. In general, it delivers an average F1-score improvement of 3.19\%.
In terms of the link loss, it delivers an average F1-score improvement of 5.12\%. For the Physics dataset, it can enhance the F1-score by 16.07\%. 
{\color{black}To evaluate the effectiveness of our proposed conductance-based augmented subgraph sampler, we replace it with the sampler in COCLEP~\cite{li2023coclep}, the previous state-of-the-art CS model. The results show that the conductance-based sampler can enhance the F1-score with an average of 3.89\%.}
Furthermore, to evaluate the effectiveness of the \transformername~architecture, we replace it with the Subg-Con model~\cite{jiao2020sub} which is a classical contrast-based self-supervised approach to pre-train the node representation. The results show that our \transformername~ significantly improves the F1-score, with an average improvement of 14.65\%.
These results collectively demonstrate the effectiveness of the modules designed in \framework.

\begin{figure}
\subfigbottomskip=-2pt
\subfigcapskip=-6pt %设置子图与子标题之间的距离
\subfigbottomskip=-3pt %设置第二行子图与第一行子图的距离，即下面的头与上面的脚的距离
    \subfigure[Ground-truth]{ 
        
        \includegraphics[width=0.11\textwidth]{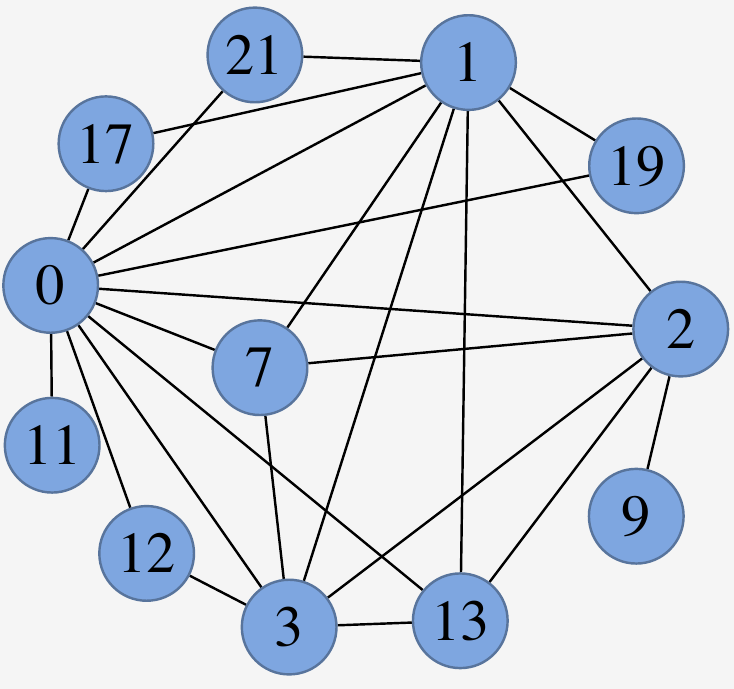}
    \vspace{-3mm}
    }
    \subfigure[TransZero]{ 
        \includegraphics[width=0.11\textwidth]{images_case_transzero.pdf}
    \vspace{-3mm}
    }
    \subfigure[QD-GNN]{ 
        \includegraphics[width=0.10\textwidth]{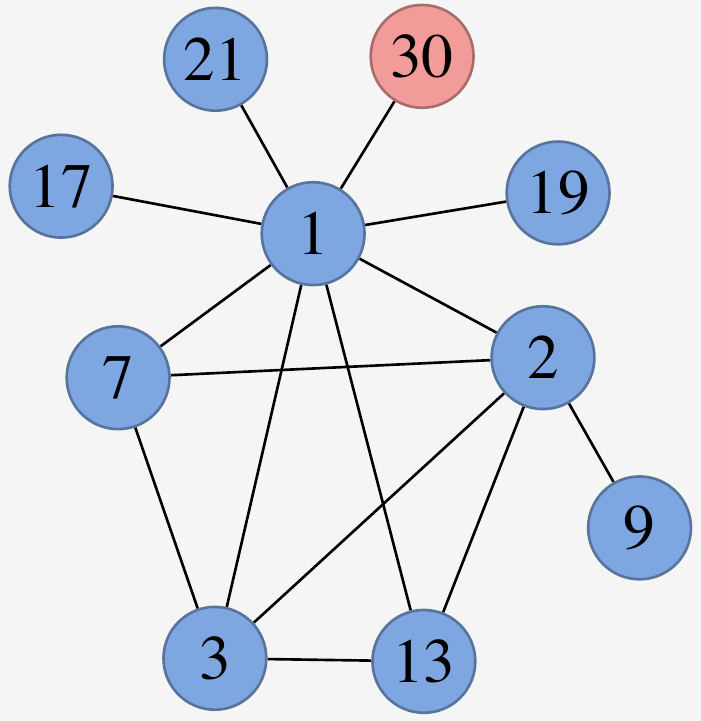}
    \vspace{-3mm}
    }
    \subfigure[COCLEP]{ 
        \includegraphics[width=0.27\textwidth]{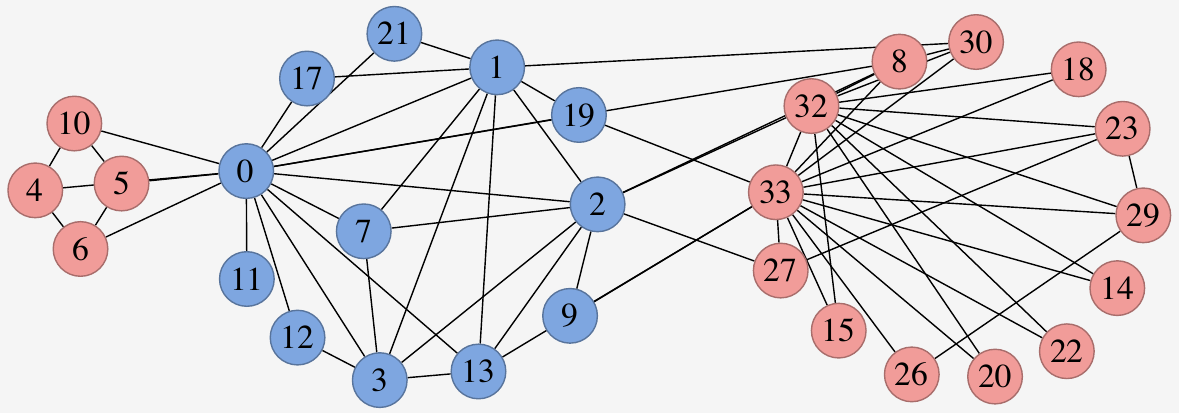}
    % \vspace{-3mm}
    }
\vspace{-2mm}
    \caption{{\color{black}Case study with query nodes \{1,9,19\}}}
    \label{fig:case_study}
\vspace{-8mm}
\end{figure}

% Karate is a real-world graph with 34 nodes and 156 edges.

{\color{black}
\noindent\textbf{Exp-11: Case study.}
We conducted a case study utilizing the real-world Zachary’s karate club network ~\cite{zachary1977information}.  We use nodes 1, 9 and 19 as query nodes. Both the ground-truth community and the results obtained from learning-based methods are depicted in Figure~\ref{fig:case_study}. The illustration demonstrates that QD-GNN fails to include some promising nodes, such as node 0, while incorporating irrelevant node  (node 30 as highlighted in red in the figure). COCLEP, on the other hand, contains numerous irrelevant nodes. In contrast, the community identified by our proposed \framework~precisely matches the ground-truth community.

}

\vspace{-2mm}
\section{Related Work}
\vspace{-1mm}
\label{sec:Relatedwork}
In this section, we give the related works about CS. Existing methods for CS can be classified into two categories: traditional CS methods and learning-based techniques. Traditional CS methods aim to identify a cohesively connected subgraph within a given graph that contains specific query nodes and satisfies given constraints. 
They model the community by pre-defined cohesive subgraph models such as \textit{k}-core~\cite{cui2014local, sozio2010community}, \textit{k}-truss~\cite{huang2014querying, akbas2017truss} and \textit{k}-edge connected component (\textit{k}-ECC)~\cite{chang2015index, hu2016querying}. Nevertheless, these approaches encounter a limitation known as \textit{structure inflexibility}.  Recently, there is a growing interest towards learning-based CS methods. ICS-GNN~\cite{gao2021ics} proposes a lightweight interactive community search model via graph neural network. QD-GNN and AQD-GNN are proposed in ~\cite{jiang2022query} for CS and attributed community search in a supervised manner. COCLEP is proposed in ~\cite{li2023coclep} for CS in a semi-supervised manner that only needs a few labels of nodes in the community rather than all labels of nodes in the ground-truth community. One parallel work is designed in~\cite{gou2023effective}. Althrough it does not use the ground-truth community information, it uses the \textit{K}-core information as labels for pretraining and predicts the \textit{K}-core community. Moreover, it selects top-\textit{K} nodes as the prediction. 

% The real-world communities may always fail to meet such strict structural constraints.

% Benefiting from the powerful approximation capacity, the learning-based method can alleviate the limitation of \textit{structure inflexibility}. 

% Our work is designed for general CS, does not rely on labels, and uses the expected score gain for the online search phase.

% \noindent \textbf{Learning-based for Community Search.}

% Recently, the interest of CS is shifted to XXX.

% \noindent \textbf{Unsupervised Learning for Graph Analytics.}  Unsupervised ML/DL techniques have shown enormous success in computer vision and natural language processing, and have also been exploited in a variety of graph-related tasks. It is introduced in ~\cite{leordeanu2012unsupervised} for graph matching. Unsupervised techniques have also had a significant impact on graph alignment~\cite{gao2021unsupervised}. An unsupervised graph neural network model is utilized in graph anomaly detection~\cite{zhao2022graph}. A self-expressive graph neural network is proposed in ~\cite{bandyopadhyay2021unsupervised} for unsupervised constrained community detection. Subg-Con is a subgraph-based graph contrastive learning for node representation~\cite{jiao2020sub}. Feature reconstruction and link reconstruction based graph representation learning are proposed in ~\cite{manessi2021graph} and ~\cite{hu2019pre}.

\vspace{-2mm}

\section{Conclusion}
\vspace{-1mm}
\label{sec:Conclusion}

In this paper, we study the problem of general community search and propose an efficient and learning-based community search framework ~\framework~that runs without using labels. It contains the offline pre-training phase and the online search phase. In the offline pre-training phase, we pre-train \transformername~which is designed specifically for the task of community search. We compute the community score without using labels by measuring the similarity of the learned representations. In the online search phase, we model the task of community identification as the task of \searchname~to free the model from using labels. We prove that the problem of \searchname~is NP-hard and APX-hard, and propose two heuristic algorithms including the \localsearch~and \globalsearch~to effectively and efficiently find promising communities. Experiments over 10 public datasets highlight the effectiveness and efficiency of \framework.

\balance
{
\bibliographystyle{ACM-Reference-Format}
\bibliography{sample}

%%% -*-BibTeX-*-
%%% Do NOT edit. File created by BibTeX with style
%%% ACM-Reference-Format-Journals [18-Jan-2012].

\begin{thebibliography}{54}

%%% ====================================================================
%%% NOTE TO THE USER: you can override these defaults by providing
%%% customized versions of any of these macros before the \bibliography
%%% command.  Each of them MUST provide its own final punctuation,
%%% except for \shownote{}, \showDOI{}, and \showURL{}.  The latter two
%%% do not use final punctuation, in order to avoid confusing it with
%%% the Web address.
%%%
%%% To suppress output of a particular field, define its macro to expand
%%% to an empty string, or better, \unskip, like this:
%%%
%%% \newcommand{\showDOI}[1]{\unskip}   % LaTeX syntax
%%%
%%% \def \showDOI #1{\unskip}           % plain TeX syntax
%%%
%%% ====================================================================

\ifx \showCODEN    \undefined \def \showCODEN     #1{\unskip}     \fi
\ifx \showDOI      \undefined \def \showDOI       #1{#1}\fi
\ifx \showISBNx    \undefined \def \showISBNx     #1{\unskip}     \fi
\ifx \showISBNxiii \undefined \def \showISBNxiii  #1{\unskip}     \fi
\ifx \showISSN     \undefined \def \showISSN      #1{\unskip}     \fi
\ifx \showLCCN     \undefined \def \showLCCN      #1{\unskip}     \fi
\ifx \shownote     \undefined \def \shownote      #1{#1}          \fi
\ifx \showarticletitle \undefined \def \showarticletitle #1{#1}   \fi
\ifx \showURL      \undefined \def \showURL       {\relax}        \fi
% The following commands are used for tagged output and should be
% invisible to TeX
\providecommand\bibfield[2]{#2}
\providecommand\bibinfo[2]{#2}
\providecommand\natexlab[1]{#1}
\providecommand\showeprint[2][]{arXiv:#2}

\bibitem[\protect\citeauthoryear{Ajtai, Koml{\'o}s, and Szemer{\'e}di}{Ajtai et~al\mbox{.}}{1983}]%
        {ajtai19830}
\bibfield{author}{\bibinfo{person}{Mikl{\'o}s Ajtai}, \bibinfo{person}{J{\'a}nos Koml{\'o}s}, {and} \bibinfo{person}{Endre Szemer{\'e}di}.} \bibinfo{year}{1983}\natexlab{}.
\newblock \showarticletitle{An 0 (n log n) sorting network}. In \bibinfo{booktitle}{\emph{Proceedings of the fifteenth annual ACM symposium on Theory of computing}}. \bibinfo{pages}{1--9}.
\newblock


\bibitem[\protect\citeauthoryear{Akbas and Zhao}{Akbas and Zhao}{2017}]%
        {akbas2017truss}
\bibfield{author}{\bibinfo{person}{Esra Akbas} {and} \bibinfo{person}{Peixiang Zhao}.} \bibinfo{year}{2017}\natexlab{}.
\newblock \showarticletitle{Truss-based community search: a truss-equivalence based indexing approach}.
\newblock \bibinfo{journal}{\emph{Proceedings of the VLDB Endowment}} \bibinfo{volume}{10}, \bibinfo{number}{11} (\bibinfo{year}{2017}), \bibinfo{pages}{1298--1309}.
\newblock


\bibitem[\protect\citeauthoryear{Akiba, Iwata, and Yoshida}{Akiba et~al\mbox{.}}{2013}]%
        {DBLP:conf/cikm/AkibaIY13}
\bibfield{author}{\bibinfo{person}{Takuya Akiba}, \bibinfo{person}{Yoichi Iwata}, {and} \bibinfo{person}{Yuichi Yoshida}.} \bibinfo{year}{2013}\natexlab{}.
\newblock \showarticletitle{Linear-time enumeration of maximal K-edge-connected subgraphs in large networks by random contraction}. In \bibinfo{booktitle}{\emph{22nd {ACM} International Conference on Information and Knowledge Management, CIKM'13, San Francisco, CA, USA, October 27 - November 1, 2013}}, \bibfield{editor}{\bibinfo{person}{Qi~He}, \bibinfo{person}{Arun Iyengar}, \bibinfo{person}{Wolfgang Nejdl}, \bibinfo{person}{Jian Pei}, {and} \bibinfo{person}{Rajeev Rastogi}} (Eds.). \bibinfo{publisher}{{ACM}}, \bibinfo{pages}{909--918}.
\newblock
\urldef\tempurl%
\url{https://doi.org/10.1145/2505515.2505751}
\showDOI{\tempurl}


\bibitem[\protect\citeauthoryear{Alon, Moshkovitz, and Safra}{Alon et~al\mbox{.}}{2006}]%
        {DBLP:journals/talg/AlonMS06}
\bibfield{author}{\bibinfo{person}{Noga Alon}, \bibinfo{person}{Dana Moshkovitz}, {and} \bibinfo{person}{Shmuel Safra}.} \bibinfo{year}{2006}\natexlab{}.
\newblock \showarticletitle{Algorithmic construction of sets for \emph{k}-restrictions}.
\newblock \bibinfo{journal}{\emph{{ACM} Trans. Algorithms}} \bibinfo{volume}{2}, \bibinfo{number}{2} (\bibinfo{year}{2006}), \bibinfo{pages}{153--177}.
\newblock
\urldef\tempurl%
\url{https://doi.org/10.1145/1150334.1150336}
\showDOI{\tempurl}


\bibitem[\protect\citeauthoryear{Alon and Yahav}{Alon and Yahav}{2021}]%
        {alon2021on}
\bibfield{author}{\bibinfo{person}{Uri Alon} {and} \bibinfo{person}{Eran Yahav}.} \bibinfo{year}{2021}\natexlab{}.
\newblock \showarticletitle{On the Bottleneck of Graph Neural Networks and its Practical Implications}. In \bibinfo{booktitle}{\emph{International Conference on Learning Representations}}.
\newblock
\urldef\tempurl%
\url{https://openreview.net/forum?id=i80OPhOCVH2}
\showURL{%
\tempurl}


\bibitem[\protect\citeauthoryear{Andersen, Chung, and Lang}{Andersen et~al\mbox{.}}{2006}]%
        {andersen2006local}
\bibfield{author}{\bibinfo{person}{Reid Andersen}, \bibinfo{person}{Fan Chung}, {and} \bibinfo{person}{Kevin Lang}.} \bibinfo{year}{2006}\natexlab{}.
\newblock \showarticletitle{Local graph partitioning using pagerank vectors}. In \bibinfo{booktitle}{\emph{2006 47th Annual IEEE Symposium on Foundations of Computer Science (FOCS'06)}}. IEEE, \bibinfo{pages}{475--486}.
\newblock


\bibitem[\protect\citeauthoryear{Arora}{Arora}{1998}]%
        {arora1998approximability}
\bibfield{author}{\bibinfo{person}{Sanjeev Arora}.} \bibinfo{year}{1998}\natexlab{}.
\newblock \showarticletitle{The approximability of NP-hard problems}. In \bibinfo{booktitle}{\emph{Proceedings of the thirtieth annual ACM symposium on Theory of computing}}. \bibinfo{pages}{337--348}.
\newblock


\bibitem[\protect\citeauthoryear{Ba, Kiros, and Hinton}{Ba et~al\mbox{.}}{2016}]%
        {ba2016layer}
\bibfield{author}{\bibinfo{person}{Jimmy~Lei Ba}, \bibinfo{person}{Jamie~Ryan Kiros}, {and} \bibinfo{person}{Geoffrey~E Hinton}.} \bibinfo{year}{2016}\natexlab{}.
\newblock \showarticletitle{Layer normalization}.
\newblock \bibinfo{journal}{\emph{arXiv preprint arXiv:1607.06450}} (\bibinfo{year}{2016}).
\newblock


\bibitem[\protect\citeauthoryear{Bi, Xu, Sun, Wang, Shen, and Cheng}{Bi et~al\mbox{.}}{2022}]%
        {bi2022company}
\bibfield{author}{\bibinfo{person}{Wendong Bi}, \bibinfo{person}{Bingbing Xu}, \bibinfo{person}{Xiaoqian Sun}, \bibinfo{person}{Zidong Wang}, \bibinfo{person}{Huawei Shen}, {and} \bibinfo{person}{Xueqi Cheng}.} \bibinfo{year}{2022}\natexlab{}.
\newblock \showarticletitle{Company-as-tribe: Company financial risk assessment on tribe-style graph with hierarchical graph neural networks}. In \bibinfo{booktitle}{\emph{Proceedings of the 28th ACM SIGKDD Conference on Knowledge Discovery and Data Mining}}. \bibinfo{pages}{2712--2720}.
\newblock


\bibitem[\protect\citeauthoryear{Cattuto, Quaggiotto, Panisson, and Averbuch}{Cattuto et~al\mbox{.}}{2013}]%
        {cattuto2013time}
\bibfield{author}{\bibinfo{person}{Ciro Cattuto}, \bibinfo{person}{Marco Quaggiotto}, \bibinfo{person}{Andr{\'e} Panisson}, {and} \bibinfo{person}{Alex Averbuch}.} \bibinfo{year}{2013}\natexlab{}.
\newblock \showarticletitle{Time-varying social networks in a graph database: a Neo4j use case}. In \bibinfo{booktitle}{\emph{First international workshop on graph data management experiences and systems}}. \bibinfo{pages}{1--6}.
\newblock


\bibitem[\protect\citeauthoryear{Chang, Lin, Qin, Yu, and Zhang}{Chang et~al\mbox{.}}{2015}]%
        {chang2015index}
\bibfield{author}{\bibinfo{person}{Lijun Chang}, \bibinfo{person}{Xuemin Lin}, \bibinfo{person}{Lu Qin}, \bibinfo{person}{Jeffrey~Xu Yu}, {and} \bibinfo{person}{Wenjie Zhang}.} \bibinfo{year}{2015}\natexlab{}.
\newblock \showarticletitle{Index-based optimal algorithms for computing steiner components with maximum connectivity}. In \bibinfo{booktitle}{\emph{Proceedings of the 2015 ACM SIGMOD International Conference on Management of Data}}. \bibinfo{pages}{459--474}.
\newblock


\bibitem[\protect\citeauthoryear{Chen, Lin, Li, Li, Zhou, and Sun}{Chen et~al\mbox{.}}{2020}]%
        {chen2020measuring}
\bibfield{author}{\bibinfo{person}{Deli Chen}, \bibinfo{person}{Yankai Lin}, \bibinfo{person}{Wei Li}, \bibinfo{person}{Peng Li}, \bibinfo{person}{Jie Zhou}, {and} \bibinfo{person}{Xu Sun}.} \bibinfo{year}{2020}\natexlab{}.
\newblock \showarticletitle{Measuring and relieving the over-smoothing problem for graph neural networks from the topological view}. In \bibinfo{booktitle}{\emph{Proceedings of the AAAI conference on artificial intelligence}}, Vol.~\bibinfo{volume}{34}. \bibinfo{pages}{3438--3445}.
\newblock


\bibitem[\protect\citeauthoryear{Chen, Gao, Li, and He}{Chen et~al\mbox{.}}{2022}]%
        {chen2022nagphormer}
\bibfield{author}{\bibinfo{person}{Jinsong Chen}, \bibinfo{person}{Kaiyuan Gao}, \bibinfo{person}{Gaichao Li}, {and} \bibinfo{person}{Kun He}.} \bibinfo{year}{2022}\natexlab{}.
\newblock \showarticletitle{NAGphormer: A tokenized graph transformer for node classification in large graphs}. In \bibinfo{booktitle}{\emph{The Eleventh International Conference on Learning Representations}}.
\newblock


\bibitem[\protect\citeauthoryear{Chen, Jiang, Zhang, and Chen}{Chen et~al\mbox{.}}{2021}]%
        {chen2021novel}
\bibfield{author}{\bibinfo{person}{Wei Chen}, \bibinfo{person}{Manrui Jiang}, \bibinfo{person}{Wei-Guo Zhang}, {and} \bibinfo{person}{Zhensong Chen}.} \bibinfo{year}{2021}\natexlab{}.
\newblock \showarticletitle{A novel graph convolutional feature based convolutional neural network for stock trend prediction}.
\newblock \bibinfo{journal}{\emph{Information Sciences}}  \bibinfo{volume}{556} (\bibinfo{year}{2021}), \bibinfo{pages}{67--94}.
\newblock


\bibitem[\protect\citeauthoryear{Chen, Zhang, and Chang}{Chen et~al\mbox{.}}{2008}]%
        {chen2008combinational}
\bibfield{author}{\bibinfo{person}{Wen-Yen Chen}, \bibinfo{person}{Dong Zhang}, {and} \bibinfo{person}{Edward~Y Chang}.} \bibinfo{year}{2008}\natexlab{}.
\newblock \showarticletitle{Combinational collaborative filtering for personalized community recommendation}. In \bibinfo{booktitle}{\emph{Proceedings of the 14th ACM SIGKDD international conference on Knowledge discovery and data mining}}. \bibinfo{pages}{115--123}.
\newblock


\bibitem[\protect\citeauthoryear{Cheng, Chen, Huang, Hsu, and Liao}{Cheng et~al\mbox{.}}{2011}]%
        {cheng2011personalized}
\bibfield{author}{\bibinfo{person}{An-Jung Cheng}, \bibinfo{person}{Yan-Ying Chen}, \bibinfo{person}{Yen-Ta Huang}, \bibinfo{person}{Winston~H Hsu}, {and} \bibinfo{person}{Hong-Yuan~Mark Liao}.} \bibinfo{year}{2011}\natexlab{}.
\newblock \showarticletitle{Personalized travel recommendation by mining people attributes from community-contributed photos}. In \bibinfo{booktitle}{\emph{Proceedings of the 19th ACM international conference on Multimedia}}. \bibinfo{pages}{83--92}.
\newblock


\bibitem[\protect\citeauthoryear{Cormen, Leiserson, Rivest, and Stein}{Cormen et~al\mbox{.}}{2022}]%
        {cormen2022introduction}
\bibfield{author}{\bibinfo{person}{Thomas~H Cormen}, \bibinfo{person}{Charles~E Leiserson}, \bibinfo{person}{Ronald~L Rivest}, {and} \bibinfo{person}{Clifford Stein}.} \bibinfo{year}{2022}\natexlab{}.
\newblock \bibinfo{booktitle}{\emph{Introduction to algorithms}}.
\newblock \bibinfo{publisher}{MIT press}.
\newblock


\bibitem[\protect\citeauthoryear{Cui, Xiao, Wang, and Wang}{Cui et~al\mbox{.}}{2014}]%
        {cui2014local}
\bibfield{author}{\bibinfo{person}{Wanyun Cui}, \bibinfo{person}{Yanghua Xiao}, \bibinfo{person}{Haixun Wang}, {and} \bibinfo{person}{Wei Wang}.} \bibinfo{year}{2014}\natexlab{}.
\newblock \showarticletitle{Local search of communities in large graphs}. In \bibinfo{booktitle}{\emph{Proceedings of the 2014 ACM SIGMOD international conference on Management of data}}. \bibinfo{pages}{991--1002}.
\newblock


\bibitem[\protect\citeauthoryear{Danon, Diaz-Guilera, Duch, and Arenas}{Danon et~al\mbox{.}}{2005}]%
        {danon2005comparing}
\bibfield{author}{\bibinfo{person}{Leon Danon}, \bibinfo{person}{Albert Diaz-Guilera}, \bibinfo{person}{Jordi Duch}, {and} \bibinfo{person}{Alex Arenas}.} \bibinfo{year}{2005}\natexlab{}.
\newblock \showarticletitle{Comparing community structure identification}.
\newblock \bibinfo{journal}{\emph{Journal of statistical mechanics: Theory and experiment}} \bibinfo{volume}{2005}, \bibinfo{number}{09} (\bibinfo{year}{2005}), \bibinfo{pages}{P09008}.
\newblock


\bibitem[\protect\citeauthoryear{Dinur and Steurer}{Dinur and Steurer}{2013}]%
        {DBLP:journals/corr/abs-1305-1979}
\bibfield{author}{\bibinfo{person}{Irit Dinur} {and} \bibinfo{person}{David Steurer}.} \bibinfo{year}{2013}\natexlab{}.
\newblock \showarticletitle{Analytical Approach to Parallel Repetition}.
\newblock \bibinfo{journal}{\emph{CoRR}}  \bibinfo{volume}{abs/1305.1979} (\bibinfo{year}{2013}).
\newblock
\showeprint[arXiv]{1305.1979}
\urldef\tempurl%
\url{http://arxiv.org/abs/1305.1979}
\showURL{%
\tempurl}


\bibitem[\protect\citeauthoryear{Fang, Huang, Qin, Zhang, Zhang, Cheng, and Lin}{Fang et~al\mbox{.}}{2020}]%
        {fang2020survey}
\bibfield{author}{\bibinfo{person}{Yixiang Fang}, \bibinfo{person}{Xin Huang}, \bibinfo{person}{Lu Qin}, \bibinfo{person}{Ying Zhang}, \bibinfo{person}{Wenjie Zhang}, \bibinfo{person}{Reynold Cheng}, {and} \bibinfo{person}{Xuemin Lin}.} \bibinfo{year}{2020}\natexlab{}.
\newblock \showarticletitle{A survey of community search over big graphs}.
\newblock \bibinfo{journal}{\emph{The VLDB Journal}}  \bibinfo{volume}{29} (\bibinfo{year}{2020}), \bibinfo{pages}{353--392}.
\newblock


\bibitem[\protect\citeauthoryear{Feige}{Feige}{1998}]%
        {feige1998threshold}
\bibfield{author}{\bibinfo{person}{Uriel Feige}.} \bibinfo{year}{1998}\natexlab{}.
\newblock \showarticletitle{A threshold of ln n for approximating set cover}.
\newblock \bibinfo{journal}{\emph{Journal of the ACM (JACM)}} \bibinfo{volume}{45}, \bibinfo{number}{4} (\bibinfo{year}{1998}), \bibinfo{pages}{634--652}.
\newblock


\bibitem[\protect\citeauthoryear{Fey and Lenssen}{Fey and Lenssen}{2019}]%
        {Fey/Lenssen/2019}
\bibfield{author}{\bibinfo{person}{Matthias Fey} {and} \bibinfo{person}{Jan~E. Lenssen}.} \bibinfo{year}{2019}\natexlab{}.
\newblock \showarticletitle{Fast Graph Representation Learning with {PyTorch Geometric}}. In \bibinfo{booktitle}{\emph{ICLR Workshop on Representation Learning on Graphs and Manifolds}}.
\newblock


\bibitem[\protect\citeauthoryear{Gao, Chen, Li, and Zhang}{Gao et~al\mbox{.}}{2021}]%
        {gao2021ics}
\bibfield{author}{\bibinfo{person}{Jun Gao}, \bibinfo{person}{Jiazun Chen}, \bibinfo{person}{Zhao Li}, {and} \bibinfo{person}{Ji Zhang}.} \bibinfo{year}{2021}\natexlab{}.
\newblock \showarticletitle{ICS-GNN: lightweight interactive community search via graph neural network}.
\newblock \bibinfo{journal}{\emph{Proceedings of the VLDB Endowment}} \bibinfo{volume}{14}, \bibinfo{number}{6} (\bibinfo{year}{2021}), \bibinfo{pages}{1006--1018}.
\newblock


\bibitem[\protect\citeauthoryear{Gou, Xu, Wu, Chen, Wang, Wu, and Ke}{Gou et~al\mbox{.}}{2023}]%
        {gou2023effective}
\bibfield{author}{\bibinfo{person}{Xiaoxuan Gou}, \bibinfo{person}{Xiaoliang Xu}, \bibinfo{person}{Xiangying Wu}, \bibinfo{person}{Runhuai Chen}, \bibinfo{person}{Yuxiang Wang}, \bibinfo{person}{Tianxing Wu}, {and} \bibinfo{person}{Xiangyu Ke}.} \bibinfo{year}{2023}\natexlab{}.
\newblock \showarticletitle{Effective and Efficient Community Search with Graph Embeddings}.
\newblock In \bibinfo{booktitle}{\emph{ECAI 2023}}. \bibinfo{publisher}{IOS Press}, \bibinfo{pages}{891--898}.
\newblock


\bibitem[\protect\citeauthoryear{Hendrycks and Gimpel}{Hendrycks and Gimpel}{2016}]%
        {hendrycks2016gaussian}
\bibfield{author}{\bibinfo{person}{Dan Hendrycks} {and} \bibinfo{person}{Kevin Gimpel}.} \bibinfo{year}{2016}\natexlab{}.
\newblock \showarticletitle{Gaussian error linear units (gelus)}.
\newblock \bibinfo{journal}{\emph{arXiv preprint arXiv:1606.08415}} (\bibinfo{year}{2016}).
\newblock


\bibitem[\protect\citeauthoryear{Hu, Wu, Cheng, Luo, and Fang}{Hu et~al\mbox{.}}{2016}]%
        {hu2016querying}
\bibfield{author}{\bibinfo{person}{Jiafeng Hu}, \bibinfo{person}{Xiaowei Wu}, \bibinfo{person}{Reynold Cheng}, \bibinfo{person}{Siqiang Luo}, {and} \bibinfo{person}{Yixiang Fang}.} \bibinfo{year}{2016}\natexlab{}.
\newblock \showarticletitle{Querying minimal steiner maximum-connected subgraphs in large graphs}. In \bibinfo{booktitle}{\emph{Proceedings of the 25th ACM International on Conference on Information and Knowledge Management}}. \bibinfo{pages}{1241--1250}.
\newblock


\bibitem[\protect\citeauthoryear{Huang, Cheng, Qin, Tian, and Yu}{Huang et~al\mbox{.}}{2014}]%
        {huang2014querying}
\bibfield{author}{\bibinfo{person}{Xin Huang}, \bibinfo{person}{Hong Cheng}, \bibinfo{person}{Lu Qin}, \bibinfo{person}{Wentao Tian}, {and} \bibinfo{person}{Jeffrey~Xu Yu}.} \bibinfo{year}{2014}\natexlab{}.
\newblock \showarticletitle{Querying k-truss community in large and dynamic graphs}. In \bibinfo{booktitle}{\emph{Proceedings of the 2014 ACM SIGMOD international conference on Management of data}}. \bibinfo{pages}{1311--1322}.
\newblock


\bibitem[\protect\citeauthoryear{Huang, Lakshmanan, Yu, and Cheng}{Huang et~al\mbox{.}}{2015}]%
        {huang2015approximate}
\bibfield{author}{\bibinfo{person}{Xin Huang}, \bibinfo{person}{Laks~VS Lakshmanan}, \bibinfo{person}{Jeffrey~Xu Yu}, {and} \bibinfo{person}{Hong Cheng}.} \bibinfo{year}{2015}\natexlab{}.
\newblock \showarticletitle{Approximate closest community search in networks}.
\newblock \bibinfo{journal}{\emph{Proceedings of the VLDB Endowment}} \bibinfo{volume}{9}, \bibinfo{number}{4} (\bibinfo{year}{2015}), \bibinfo{pages}{276--287}.
\newblock


\bibitem[\protect\citeauthoryear{Jiang, Rong, Cheng, Huang, Zhao, and Huang}{Jiang et~al\mbox{.}}{2022}]%
        {jiang2022query}
\bibfield{author}{\bibinfo{person}{Yuli Jiang}, \bibinfo{person}{Yu Rong}, \bibinfo{person}{Hong Cheng}, \bibinfo{person}{Xin Huang}, \bibinfo{person}{Kangfei Zhao}, {and} \bibinfo{person}{Junzhou Huang}.} \bibinfo{year}{2022}\natexlab{}.
\newblock \showarticletitle{Query driven-graph neural networks for community search: from non-attributed, attributed, to interactive attributed}.
\newblock \bibinfo{journal}{\emph{Proceedings of the VLDB Endowment}} \bibinfo{volume}{15}, \bibinfo{number}{6} (\bibinfo{year}{2022}), \bibinfo{pages}{1243--1255}.
\newblock


\bibitem[\protect\citeauthoryear{Jiao, Xiong, Zhang, Zhang, Zhang, and Zhu}{Jiao et~al\mbox{.}}{2020}]%
        {jiao2020sub}
\bibfield{author}{\bibinfo{person}{Yizhu Jiao}, \bibinfo{person}{Yun Xiong}, \bibinfo{person}{Jiawei Zhang}, \bibinfo{person}{Yao Zhang}, \bibinfo{person}{Tianqi Zhang}, {and} \bibinfo{person}{Yangyong Zhu}.} \bibinfo{year}{2020}\natexlab{}.
\newblock \showarticletitle{Sub-graph contrast for scalable self-supervised graph representation learning}. In \bibinfo{booktitle}{\emph{2020 IEEE international conference on data mining (ICDM)}}. IEEE, \bibinfo{pages}{222--231}.
\newblock


\bibitem[\protect\citeauthoryear{Kim, Luo, Cong, and Yu}{Kim et~al\mbox{.}}{2022}]%
        {kim2022dmcs}
\bibfield{author}{\bibinfo{person}{Junghoon Kim}, \bibinfo{person}{Siqiang Luo}, \bibinfo{person}{Gao Cong}, {and} \bibinfo{person}{Wenyuan Yu}.} \bibinfo{year}{2022}\natexlab{}.
\newblock \showarticletitle{DMCS: Density modularity based community search}. In \bibinfo{booktitle}{\emph{Proceedings of the 2022 International Conference on Management of Data}}. \bibinfo{pages}{889--903}.
\newblock


\bibitem[\protect\citeauthoryear{Li, Luo, Zhao, Shan, Wang, and Qin}{Li et~al\mbox{.}}{2023}]%
        {li2023coclep}
\bibfield{author}{\bibinfo{person}{Ling Li}, \bibinfo{person}{Siqiang Luo}, \bibinfo{person}{Yuhai Zhao}, \bibinfo{person}{Caihua Shan}, \bibinfo{person}{Zhengkui Wang}, {and} \bibinfo{person}{Lu Qin}.} \bibinfo{year}{2023}\natexlab{}.
\newblock \showarticletitle{COCLEP: Contrastive Learning-based Semi-Supervised Community Search}.
\newblock \bibinfo{journal}{\emph{IEEE 39th ICDE}} (\bibinfo{year}{2023}).
\newblock


\bibitem[\protect\citeauthoryear{Li, Hui, Zhang, Huang, Wang, Tian, Zhang, Gao, and Tang}{Li et~al\mbox{.}}{2021}]%
        {li2021happens}
\bibfield{author}{\bibinfo{person}{Zhao Li}, \bibinfo{person}{Pengrui Hui}, \bibinfo{person}{Peng Zhang}, \bibinfo{person}{Jiaming Huang}, \bibinfo{person}{Biao Wang}, \bibinfo{person}{Ling Tian}, \bibinfo{person}{Ji Zhang}, \bibinfo{person}{Jianliang Gao}, {and} \bibinfo{person}{Xing Tang}.} \bibinfo{year}{2021}\natexlab{}.
\newblock \showarticletitle{What happens behind the scene? Towards fraud community detection in e-commerce from online to offline}. In \bibinfo{booktitle}{\emph{Companion Proceedings of the Web Conference 2021}}. \bibinfo{pages}{105--113}.
\newblock


\bibitem[\protect\citeauthoryear{Liu, Jin, Pan, Zhou, Zheng, Xia, and Philip}{Liu et~al\mbox{.}}{2022}]%
        {liu2022graph}
\bibfield{author}{\bibinfo{person}{Yixin Liu}, \bibinfo{person}{Ming Jin}, \bibinfo{person}{Shirui Pan}, \bibinfo{person}{Chuan Zhou}, \bibinfo{person}{Yu Zheng}, \bibinfo{person}{Feng Xia}, {and} \bibinfo{person}{S~Yu Philip}.} \bibinfo{year}{2022}\natexlab{}.
\newblock \showarticletitle{Graph self-supervised learning: A survey}.
\newblock \bibinfo{journal}{\emph{IEEE Transactions on Knowledge and Data Engineering}} \bibinfo{volume}{35}, \bibinfo{number}{6} (\bibinfo{year}{2022}), \bibinfo{pages}{5879--5900}.
\newblock


\bibitem[\protect\citeauthoryear{Qi, Balem, Faloutsos, Klein-Seetharaman, and Bar-Joseph}{Qi et~al\mbox{.}}{2008}]%
        {qi2008protein}
\bibfield{author}{\bibinfo{person}{Yanjun Qi}, \bibinfo{person}{Fernanda Balem}, \bibinfo{person}{Christos Faloutsos}, \bibinfo{person}{Judith Klein-Seetharaman}, {and} \bibinfo{person}{Ziv Bar-Joseph}.} \bibinfo{year}{2008}\natexlab{}.
\newblock \showarticletitle{Protein complex identification by supervised graph local clustering}.
\newblock \bibinfo{journal}{\emph{Bioinformatics}} \bibinfo{volume}{24}, \bibinfo{number}{13} (\bibinfo{year}{2008}), \bibinfo{pages}{i250--i268}.
\newblock


\bibitem[\protect\citeauthoryear{Qi, Dong, Fan, Ge, Zhang, Ma, and Yi}{Qi et~al\mbox{.}}{2023}]%
        {qi2023recon}
\bibfield{author}{\bibinfo{person}{Zekun Qi}, \bibinfo{person}{Runpei Dong}, \bibinfo{person}{Guofan Fan}, \bibinfo{person}{Zheng Ge}, \bibinfo{person}{Xiangyu Zhang}, \bibinfo{person}{Kaisheng Ma}, {and} \bibinfo{person}{Li Yi}.} \bibinfo{year}{2023}\natexlab{}.
\newblock \showarticletitle{Contrast with Reconstruct: Contrastive 3D Representation Learning Guided by Generative Pretraining}. In \bibinfo{booktitle}{\emph{International Conference on Machine Learning (ICML)}}.
\newblock


\bibitem[\protect\citeauthoryear{Sankar and Chandra}{Sankar and Chandra}{2022}]%
        {sankar2022sitemotif}
\bibfield{author}{\bibinfo{person}{Santhosh Sankar} {and} \bibinfo{person}{Nagasuma Chandra}.} \bibinfo{year}{2022}\natexlab{}.
\newblock \showarticletitle{SiteMotif: A graph-based algorithm for deriving structural motifs in Protein Ligand binding sites}.
\newblock \bibinfo{journal}{\emph{PLoS Computational Biology}} \bibinfo{volume}{18}, \bibinfo{number}{2} (\bibinfo{year}{2022}), \bibinfo{pages}{e1009901}.
\newblock


\bibitem[\protect\citeauthoryear{Sasaki et~al\mbox{.}}{Sasaki et~al\mbox{.}}{2007}]%
        {sasaki2007truth}
\bibfield{author}{\bibinfo{person}{Yutaka Sasaki} {et~al\mbox{.}}} \bibinfo{year}{2007}\natexlab{}.
\newblock \showarticletitle{The truth of the F-measure}.
\newblock \bibinfo{journal}{\emph{Teach tutor mater}} \bibinfo{volume}{1}, \bibinfo{number}{5} (\bibinfo{year}{2007}), \bibinfo{pages}{1--5}.
\newblock


\bibitem[\protect\citeauthoryear{Schlosser and Wagner}{Schlosser and Wagner}{2004}]%
        {schlosser2004modularity}
\bibfield{author}{\bibinfo{person}{Gerhard Schlosser} {and} \bibinfo{person}{G{\"u}nter~P Wagner}.} \bibinfo{year}{2004}\natexlab{}.
\newblock \showarticletitle{Modularity in development and evolution}.
\newblock  (\bibinfo{year}{2004}).
\newblock


\bibitem[\protect\citeauthoryear{Schroff, Kalenichenko, and Philbin}{Schroff et~al\mbox{.}}{2015}]%
        {schroff2015facenet}
\bibfield{author}{\bibinfo{person}{Florian Schroff}, \bibinfo{person}{Dmitry Kalenichenko}, {and} \bibinfo{person}{James Philbin}.} \bibinfo{year}{2015}\natexlab{}.
\newblock \showarticletitle{Facenet: A unified embedding for face recognition and clustering}. In \bibinfo{booktitle}{\emph{Proceedings of the IEEE conference on computer vision and pattern recognition}}. \bibinfo{pages}{815--823}.
\newblock


\bibitem[\protect\citeauthoryear{Sozio and Gionis}{Sozio and Gionis}{2010}]%
        {sozio2010community}
\bibfield{author}{\bibinfo{person}{Mauro Sozio} {and} \bibinfo{person}{Aristides Gionis}.} \bibinfo{year}{2010}\natexlab{}.
\newblock \showarticletitle{The community-search problem and how to plan a successful cocktail party}. In \bibinfo{booktitle}{\emph{Proceedings of the 16th ACM SIGKDD international conference on Knowledge discovery and data mining}}. \bibinfo{pages}{939--948}.
\newblock


\bibitem[\protect\citeauthoryear{Tang and Liu}{Tang and Liu}{2010}]%
        {tang2010graph}
\bibfield{author}{\bibinfo{person}{Lei Tang} {and} \bibinfo{person}{Huan Liu}.} \bibinfo{year}{2010}\natexlab{}.
\newblock \showarticletitle{Graph mining applications to social network analysis}.
\newblock \bibinfo{journal}{\emph{Managing and mining graph data}} (\bibinfo{year}{2010}), \bibinfo{pages}{487--513}.
\newblock


\bibitem[\protect\citeauthoryear{Wang, You, Li, Zheng, and Huang}{Wang et~al\mbox{.}}{2020}]%
        {wang2020gcncda}
\bibfield{author}{\bibinfo{person}{Lei Wang}, \bibinfo{person}{Zhu-Hong You}, \bibinfo{person}{Yang-Ming Li}, \bibinfo{person}{Kai Zheng}, {and} \bibinfo{person}{Yu-An Huang}.} \bibinfo{year}{2020}\natexlab{}.
\newblock \showarticletitle{GCNCDA: a new method for predicting circRNA-disease associations based on graph convolutional network algorithm}.
\newblock \bibinfo{journal}{\emph{PLOS Computational Biology}} \bibinfo{volume}{16}, \bibinfo{number}{5} (\bibinfo{year}{2020}), \bibinfo{pages}{e1007568}.
\newblock


\bibitem[\protect\citeauthoryear{Wu, Wang, Feng, He, Chen, Lian, and Xie}{Wu et~al\mbox{.}}{2021}]%
        {wu2021self}
\bibfield{author}{\bibinfo{person}{Jiancan Wu}, \bibinfo{person}{Xiang Wang}, \bibinfo{person}{Fuli Feng}, \bibinfo{person}{Xiangnan He}, \bibinfo{person}{Liang Chen}, \bibinfo{person}{Jianxun Lian}, {and} \bibinfo{person}{Xing Xie}.} \bibinfo{year}{2021}\natexlab{}.
\newblock \showarticletitle{Self-supervised graph learning for recommendation}. In \bibinfo{booktitle}{\emph{Proceedings of the 44th international ACM SIGIR conference on research and development in information retrieval}}. \bibinfo{pages}{726--735}.
\newblock


\bibitem[\protect\citeauthoryear{Wu, Jin, Li, and Zhang}{Wu et~al\mbox{.}}{2015}]%
        {wu2015robust}
\bibfield{author}{\bibinfo{person}{Yubao Wu}, \bibinfo{person}{Ruoming Jin}, \bibinfo{person}{Jing Li}, {and} \bibinfo{person}{Xiang Zhang}.} \bibinfo{year}{2015}\natexlab{}.
\newblock \showarticletitle{Robust local community detection: on free rider effect and its elimination}.
\newblock \bibinfo{journal}{\emph{Proceedings of the VLDB Endowment}} \bibinfo{volume}{8}, \bibinfo{number}{7} (\bibinfo{year}{2015}), \bibinfo{pages}{798--809}.
\newblock


\bibitem[\protect\citeauthoryear{Yang and Leskovec}{Yang and Leskovec}{2012}]%
        {yang2012defining}
\bibfield{author}{\bibinfo{person}{Jaewon Yang} {and} \bibinfo{person}{Jure Leskovec}.} \bibinfo{year}{2012}\natexlab{}.
\newblock \showarticletitle{Defining and evaluating network communities based on ground-truth}. In \bibinfo{booktitle}{\emph{Proceedings of the ACM SIGKDD Workshop on Mining Data Semantics}}. \bibinfo{pages}{1--8}.
\newblock


\bibitem[\protect\citeauthoryear{Yang, Zhang, Zhou, Wang, Sun, Zhong, Fang, Yu, and Qi}{Yang et~al\mbox{.}}{2021}]%
        {yang2021financial}
\bibfield{author}{\bibinfo{person}{Shuo Yang}, \bibinfo{person}{Zhiqiang Zhang}, \bibinfo{person}{Jun Zhou}, \bibinfo{person}{Yang Wang}, \bibinfo{person}{Wang Sun}, \bibinfo{person}{Xingyu Zhong}, \bibinfo{person}{Yanming Fang}, \bibinfo{person}{Quan Yu}, {and} \bibinfo{person}{Yuan Qi}.} \bibinfo{year}{2021}\natexlab{}.
\newblock \showarticletitle{Financial risk analysis for SMEs with graph-based supply chain mining}. In \bibinfo{booktitle}{\emph{Proceedings of the Twenty-Ninth International Conference on International Joint Conferences on Artificial Intelligence}}. \bibinfo{pages}{4661--4667}.
\newblock


\bibitem[\protect\citeauthoryear{Ying, Cai, Luo, Zheng, Ke, He, Shen, and Liu}{Ying et~al\mbox{.}}{2021}]%
        {ying2021transformers}
\bibfield{author}{\bibinfo{person}{Chengxuan Ying}, \bibinfo{person}{Tianle Cai}, \bibinfo{person}{Shengjie Luo}, \bibinfo{person}{Shuxin Zheng}, \bibinfo{person}{Guolin Ke}, \bibinfo{person}{Di He}, \bibinfo{person}{Yanming Shen}, {and} \bibinfo{person}{Tie-Yan Liu}.} \bibinfo{year}{2021}\natexlab{}.
\newblock \showarticletitle{Do transformers really perform badly for graph representation?}
\newblock \bibinfo{journal}{\emph{Advances in Neural Information Processing Systems}}  \bibinfo{volume}{34} (\bibinfo{year}{2021}), \bibinfo{pages}{28877--28888}.
\newblock


\bibitem[\protect\citeauthoryear{Zachary}{Zachary}{1977}]%
        {zachary1977information}
\bibfield{author}{\bibinfo{person}{Wayne~W Zachary}.} \bibinfo{year}{1977}\natexlab{}.
\newblock \showarticletitle{An information flow model for conflict and fission in small groups}.
\newblock \bibinfo{journal}{\emph{Journal of anthropological research}} \bibinfo{volume}{33}, \bibinfo{number}{4} (\bibinfo{year}{1977}), \bibinfo{pages}{452--473}.
\newblock


\bibitem[\protect\citeauthoryear{Zhang, Zhou, Yildirim, Alcorn, He, Davulcu, and Tong}{Zhang et~al\mbox{.}}{2017}]%
        {zhang2017hidden}
\bibfield{author}{\bibinfo{person}{Si Zhang}, \bibinfo{person}{Dawei Zhou}, \bibinfo{person}{Mehmet~Yigit Yildirim}, \bibinfo{person}{Scott Alcorn}, \bibinfo{person}{Jingrui He}, \bibinfo{person}{Hasan Davulcu}, {and} \bibinfo{person}{Hanghang Tong}.} \bibinfo{year}{2017}\natexlab{}.
\newblock \showarticletitle{Hidden: hierarchical dense subgraph detection with application to financial fraud detection}. In \bibinfo{booktitle}{\emph{Proceedings of the 2017 SIAM international conference on data mining}}. SIAM, \bibinfo{pages}{570--578}.
\newblock


\bibitem[\protect\citeauthoryear{Zhang, Xiong, Ye, Liu, Wang, Zhu, and Yu}{Zhang et~al\mbox{.}}{2020}]%
        {zhang2020seal}
\bibfield{author}{\bibinfo{person}{Yao Zhang}, \bibinfo{person}{Yun Xiong}, \bibinfo{person}{Yun Ye}, \bibinfo{person}{Tengfei Liu}, \bibinfo{person}{Weiqiang Wang}, \bibinfo{person}{Yangyong Zhu}, {and} \bibinfo{person}{Philip~S Yu}.} \bibinfo{year}{2020}\natexlab{}.
\newblock \showarticletitle{SEAL: Learning heuristics for community detection with generative adversarial networks}. In \bibinfo{booktitle}{\emph{Proceedings of the 26th ACM SIGKDD International Conference on Knowledge Discovery \& Data Mining}}. \bibinfo{pages}{1103--1113}.
\newblock


\bibitem[\protect\citeauthoryear{Zhang, Liu, Wang, Lu, and Lee}{Zhang et~al\mbox{.}}{2021}]%
        {zhang2021motif}
\bibfield{author}{\bibinfo{person}{Zaixi Zhang}, \bibinfo{person}{Qi Liu}, \bibinfo{person}{Hao Wang}, \bibinfo{person}{Chengqiang Lu}, {and} \bibinfo{person}{Chee-Kong Lee}.} \bibinfo{year}{2021}\natexlab{}.
\newblock \showarticletitle{Motif-based graph self-supervised learning for molecular property prediction}.
\newblock \bibinfo{journal}{\emph{Advances in Neural Information Processing Systems}}  \bibinfo{volume}{34} (\bibinfo{year}{2021}), \bibinfo{pages}{15870--15882}.
\newblock


\bibitem[\protect\citeauthoryear{Zhao, Li, Wen, Wang, Liu, Sun, Xie, and Ye}{Zhao et~al\mbox{.}}{2021}]%
        {zhao2021gophormer}
\bibfield{author}{\bibinfo{person}{Jianan Zhao}, \bibinfo{person}{Chaozhuo Li}, \bibinfo{person}{Qianlong Wen}, \bibinfo{person}{Yiqi Wang}, \bibinfo{person}{Yuming Liu}, \bibinfo{person}{Hao Sun}, \bibinfo{person}{Xing Xie}, {and} \bibinfo{person}{Yanfang Ye}.} \bibinfo{year}{2021}\natexlab{}.
\newblock \showarticletitle{Gophormer: Ego-graph transformer for node classification}.
\newblock \bibinfo{journal}{\emph{arXiv preprint arXiv:2110.13094}} (\bibinfo{year}{2021}).
\newblock


\end{thebibliography}
}

\end{document}